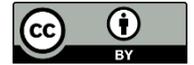

# Thermoelectric properties of Sm-doped BiCuSeO oxyselenides fabricated by two-step reactive sintering[§]

Andrei Novitskii,[a,b,*] Illia Serhiienko,[a,c,d] Sergey Novikov,[b] Kirill Kuskov,[a] Daria Pankratova,[a,‡] Tatyana Sviridova,[a] Andrei Voronin,[a] Aleksei Bogach,[e] Elena Skryleva,[a] Yuriy Parkhomenko,[a,f] Alexander Burkov,[b] Takao Mori,[c,d] Vladimir Khovaylo[a]

***Abstract:*** Among layered oxygen-containing compounds, BiCuSeO is one of the most promising candidates for thermoelectric applications due to its intrinsically low thermal conductivity and good thermal stability. However, the rather poor electrical conductivity of pristine BiCuSeO hinders its potential. Further enhancement of the thermoelectric performance by single doping at the Bi site is limited mainly due to a dramatic decrease in carrier mobility. Thus, new strategies, such as dual doping or doping with variable-valence elements seem to be promising. Along with that, the development of a fast and scalable synthesis route is essential for the industrial-scale fabrication of thermoelectric materials. Hence, in this paper, $Bi_{1-x}Sm_xCuSeO$ samples ($0 \leq x \leq 0.08$) have been synthesized with a simple and scalable reactive sintering process. For comparison, $Bi_{1-x}Sm_xCuSeO$ oxyselenides were also obtained by the conventional solid-state route. Our results highlight that Sm for Bi substitution increases the electrical conductivity by 1.5 – 2 times and decreases the Seebeck coefficient by ~1.4 times at 873 K for both series. Overall, considering the increase of lattice thermal conductivity upon doping and not optimized power factor, the figure of merit $zT$ is reduced upon doping.

## Introduction

Since the work by Vancliff Johnson and Wolfgang Jeitschko in 1974, the 1111 phases with ZrCuSiAs structure type have attracted considerable attention in the materials science community due to their quasi-two-dimensional crystal structure and nontrivial system of interatomic bonds [1–3]. These ZrCuSiAs-like phases formulated as $RTChQ$ (where $R$ = Bi, Ba, Sr or rare-earth elements; $T$ = late transition metals; $Ch$ = P, Sb, Si, Ge or chalcogens; $Q$ = O, F, H, As) typically crystallize in a layered structure composed of alternating conductive [$T_2Ch_2$] and charge reservoir [$R_2Q_2$] layers [2]. The 1111 phases were thoroughly studied as new potential high-temperature superconductors and as transparent *p*-type semiconductors for optoelectronic applications [4–7]. One of the first studies on the electrical transport properties of BiCuSeO oxyselenides was barely noticed in 1997 [8]. However, almost a decade later it was clearly shown by two groups led by Nita Dragoe and Anthony V. Powell that the BiCu$Ch$O-based ($Ch$ = Se or Te) oxychalcogenides could be promising as *p*-type thermoelectric (TE) material mainly due to their intrinsically low thermal conductivity (< 1.5 W m$^{-1}$ K$^{-1}$ at room temperature) and moderate electrical transport properties [9,10]. Indeed, a few years later a high $zT$ of ~1.5 at ~900 K was achieved by several groups for BiCuSeO doped with $Ba_{Bi}$ or $Ca_{Bi}$ and $Pb_{Bi}$, which makes them among the best performing oxygen-containing thermoelectric materials [11–14]. Here, $zT$ is the dimensionless figure of merit related to the efficiency of heat to electrical energy conversion, which is defined as $zT = \alpha^2 \sigma T/\kappa$, where $\alpha$, $\sigma$, $T$, $\kappa$ are the Seebeck coefficient, the electrical conductivity, the absolute temperature, and the thermal conductivity, respectively.

In general, good TE material should have high $\alpha$ and $\sigma$ along with low $\kappa$ [15–18]. However, it is usually difficult to achieve since the charge carriers and phonons transports are strongly coupled with each other [19]. BiCuSeO oxyselenides exhibit intrinsically low thermal conductivity as mentioned before and thus, the main strategies for the BiCuSeO thermoelectric performance enhancement are realized through the tailoring of the electrical transport properties, e.g., (i) charge carrier concentration optimization *via* doping at Bi site [20–22]; (ii) band structure engineering *via* isovalent substitution of Bi or Se [23,24]; (iii) defect engineering [25,26]. In most cases, higher $zT$ is achieved when several strategies are applied at the same time [27–31]. In this regard doping BiCuSeO with variable-valence elements, such as Eu, Yb, or Sm, which prefer a divalent state, can be considered a promising strategy for the $zT$ optimization *via* both band structure tuning and carrier density increase due to hole introduction [32–34].

In addition to the above, all the thermoelectric properties are strongly coupled with the multiscale microstructures of the material. It is why the huge attention of the scientific community is also devoted to the development and application of advanced material synthesis and processing techniques [35]. Usually, the nonequilibrium routes, such as melt spinning (MS) [36], self-propagating high-temperature synthesis (SHS) [37], mechanical alloying (MA) [38], etc., are used to improve the figure of merit *via* achieving rich hierarchical micro- or nanostructures [39–41]. Nevertheless, all of these methods require a subsequent sintering step to obtain a bulk material. For this purpose, the spark plasma sintering (SPS) technique (also known as field-assisted or pulsed-electric current sintering) is widely used due to its outstanding energy and time efficacy [42]. However, the SPS can be used not only as a densification step but also as a chemical reaction stage (so-called reactive SPS, RSPS, or RS), which, in contrast, allows one to obtain the material in bulk directly from the precursors [43]. In comparison with the conventional solid-state reaction (SSR) approach, RSPS can be considered a straightforward and potentially scalable synthesis since some materials can be successfully fabricated by the RSPS in a matter of minutes [43]. However, all the advantages of the RSPS should be considered with caution. The spark plasma process is not yet fully understood and there are still some debates over the presence of sparks and plasmas in the process, the relationship between a material's electrical conductivity and temperature evolution during the sintering, the thermodynamics, and kinetics of the RSPS processes, etc. For more details see reviews by O. Guillon *et al.*, and A.S. Mukasyan *et al.* [42,43]. Nevertheless, the RSPS approach was already reported to be appropriate for many thermoelectric systems [44–47]. Moreover, it was recently reported by our group that the BiCuSeO oxyselenide in a bulk form obtained by reactive sintering exhibits a comparable $zT$ value to that of BiCuSeO fabricated by SSR [48].

---

[a] National University of Science and Technology MISIS, 119049 Moscow, Russian Federation
[b] Ioffe Institute, 194021 St. Petersburg, Russian Federation
[c] National Institute for Materials Science (NIMS), International Center for Materials Nanoarchitectonics (WPI-MANA), 305-0044 Tsukuba, Japan
[d] University of Tsukuba, Graduate School of Pure and Applied Sciences, 305-8671 Tsukuba, Japan
[e] Prokhorov General Physics Institute of the Russian Academy of Sciences, 119991 Moscow, Russian Federation
[f] JSC "Giredmet", 111524 Moscow, Russian Federation
[*] E-mail: novitskiy@misis.ru
[‡] Present address: Luleå University of Technology, 97187 Luleå, Sweden
[§] Electronic Supplementary Information (ESI) available

Thus, in this report, we combine the investigation of the influence of BiCuSeO doping with variable-valence Sm at the Bi site on the thermoelectric properties and a comparison of two different fabrication routes, namely the so-called two-step reactive sintering and a conventional two-step solid-state reaction route. It is worth noting that the thermoelectric properties of Sm-doped BiCuSeO prepared by MA and SSR were already reported by B. Feng *et al.* and H. Kang *et al.*, respectively [33,34]. The previously reported results are considered and thoroughly discussed.

## Experimental details

The bulk samples with the nominal chemical composition of $Bi_{1-x}Sm_xCuSeO$ ($x$ = 0, 0.02, 0.04, 0.06, and 0.08) were synthesized by means of two previously mentioned methods: (i) two-step reactive sintering (samples were labeled as RS) and (ii) two-step solid-state reaction combined with ball milling (BM) and spark plasma sintering (samples were labeled as SS). For both routes, fine commercial powders of $Bi_2O_3$ (99.5%, Reachem), $Sm_2O_3$ (99.99%, Shin-Etsu Chemical), Bi (≥ 99.95%, Component Reaktiv), Se (99.90%, Reachem), and Cu (≥ 99.5%, Rushim) were used as precursors (see Table S1). While the preparation of the oxyselenide phase in the powder form *via* the SSR route requires two steps of long-term annealing in a quartz tube along with intermediate BM [49], the RS technique involves only a short step of mixing precursors followed by RS and BM [48]. Then, in both cases, resultant powders were placed in Ø10 mm graphite dies and densified by SPS (Labox 650, Sinter-Land, Japan) under a uniaxial pressure of 50 MPa under Ar atmosphere at 973 K for 5 min (heating rate of 50 K/min and cooling rate of 20 K/min). It should be noted that the RS and SS samples were fabricated by using exactly the same conditions and regimes as reported in our previous reports, respectively [48,49].

X-ray diffraction (XRD) patterns were collected at room temperature using a DRON-3 diffractometer (IC Bourevestnik, Russia) equipped with a $CuK_\alpha$ ($\lambda$ = 1.54178 Å) X-ray tube. In order to quantitatively estimate the degree of preferential grain orientation in the obtained polycrystalline samples, the XRD patterns of both powders and bulks after sintering were analyzed and refined by the Rietveld method with the self-developed software package [50]. A two-dimensional projection of the orientation distribution function, the so-called pole density (or pole distribution) was calculated from the patterns using the following formulas [51–53]:

$$I_{HKL} = kP_{hkl}M_{hkl},$$
$$I^0_{HKL} = k_0 P_0 M_{hkl}. \qquad (1)$$

Here, $I_{HKL}$ and $I^0_{HKL}$ are the integrated intensity of the HKL line corresponding to the textured and non-textured sample, respectively, $k$ is the coefficients of proportionality determined by intensity factors and experimental conditions, $P$ is the pole density, which indicates a number of normals to the set of $\{hkl\}$ planes per unit area of the sphere of the reciprocal lattice node, $M_{hkl}$ is the multiplicity factor. The ratio between $I_{HKL}$ and $I^0_{HKL}$ is equal to the ratio between the volumes of crystallites in reflection for the textured and non-textured samples. This ratio, $\Phi_{hkl}$, is called the relative pole density (or pole distribution function) and considering that the average value of the pole density $\overline{P} = \sum_n P_{hkl}/n = P_0 = 1$ when $n \to \infty$, can be calculated by the following expression [51]:

$$\Phi_{hkl} = \frac{\dfrac{I_{HKL} \cdot n}{I^0_{HKL}}}{\sum_n \dfrac{I_{HKL}}{I^0_{HKL}}}, \qquad (2)$$

where $n$ is the number of lines for which the calculation is carried out. $\Phi_{hkl}$ shows how the probability for $\{hkl\}$ to be parallel to the sample surface in a textured sample differs from that in a non-textured one ($\Phi_{hkl} \geq 0$). Thus, $\Phi_{hkl} = 0$ means that there are no crystallites with $\{hkl\}$ planes, $\Phi_{hkl} = 1$ means that there is no preferential crystallites orientation in the sample and, for example, $\Phi_{hkl} = 2$ means that there are twice as many crystallites in the textured sample, the $\{hkl\}$ planes of which are oriented parallel to the sample surface. For the $\Phi_{hkl}$ calculation, only reflections of the same order from the $\{hkl\}$ planes were taken into account, although reflections of different orders were also recorded in the diffraction patterns.

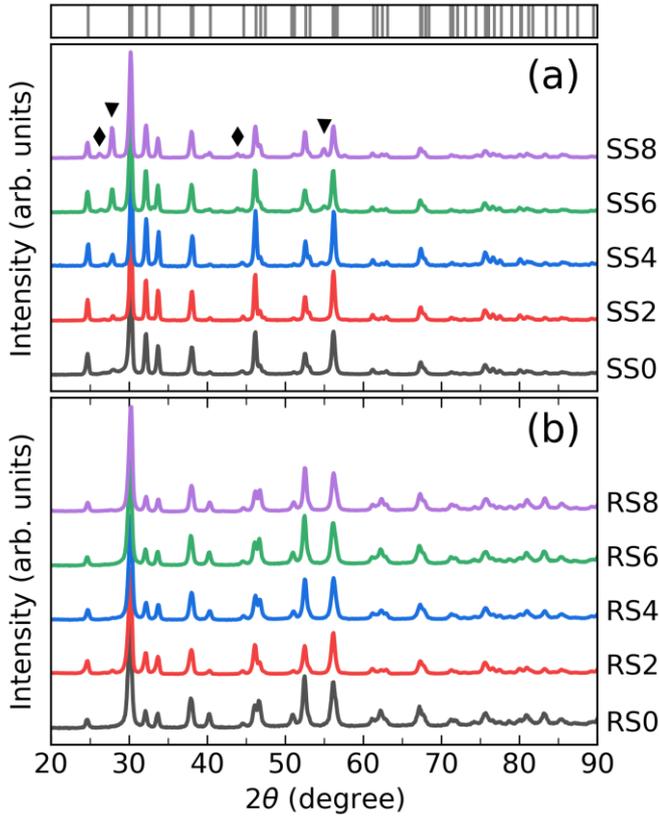

**Figure 1.** X-ray diffraction patterns for $Bi_{1-x}Sm_xCuSeO$ ($x = 0 - 0.08$) bulks prepared by both (a) SS and (b) RS techniques. The Bragg's reflections for BiCuSeO phase are indicated by grey ticks on the top part of the figure. $Bi_2O_3$ and $Cu_{2-x}Se$ secondary phases are indicated by black solid triangle (▼) and thin diamond (♦) symbols, respectively.

Microstructures and elemental composition of sintered samples were examined by scanning electron microscopy (SEM) and energy-dispersive X-ray spectroscopy (EDS) on Vega 3SB scanning electron microscope (Tescan, Czech Republic) in conjunction with an EDS detector (x-act, Oxford Instruments, UK). The accuracy of the EDS analysis is limited and affected by many factors, including the chemical composition of the sample itself, near edge absorption effects, parameters of the detector, etc. [54]. Thus, in order to obtain reasonable results, we have carried out not less than 30 point-type and 10 area-type analyses per sample. Overall, the uncertainty of the obtained atomic fraction of Sm was estimated to be $10 - 15\%$. X-ray photoelectron spectroscopic (XPS) measurements were performed in a UHV system by using PHI5000VersaProbeII spectrometer (ULVAC-PHI, Japan). A monochromatic Al K$\alpha$ radiation source ($hv = 1486.6$ eV) was employed with a power of 50 W. The spot size was 200 μm. The spectrometer scale of the binding energy was calibrated by Au $4f^7$ (83.96 eV) and Cu $2p^3$ (932.62 eV). The peaks were deconvoluted by means of an open-source Python package LG4X [55] in order to resolve the separate constituents.

The Seebeck coefficient, $\alpha$, and the electrical conductivity, $\sigma$, were simultaneously measured in the radial direction (perpendicular to the SPS pressure direction) of bar-shaped samples with dimensions of 10 mm × 2 mm × 2 mm by the standard four-probe and differential methods, respectively. The measurements were performed under a helium atmosphere using the homemade system developed at Ioffe Institute [56]. In order to assess the thermal stability during the measurement and give repeatable data the electrical transport properties measurements were carried out on heating and cooling cycles. The thermal diffusivity, $\chi$, was measured in the axial direction (parallel to the SPS pressure direction) of a disk-shaped sample of Ø 10 mm × 1 mm using an LFA 467 HyperFlash (Netzsch, Germany). The thermal conductivity, $\kappa$, was calculated by $\kappa = \chi \cdot C_p \cdot d$, where $C_p$ is the specific heat capacity calculated by the Debye model [57], and $d$ is the volume density measured by the Archimedes method. Hall constant at room temperature was calculated from the angular dependency of the Hall resistivity $\rho_H(\varphi)$ measured using a laboratory-made installation (Cryotel, Russia) by the so-called stepwise rotation technique with a fixed magnetic field of 8 T, applied perpendicular to the rotation axis. $\rho_H(\varphi)$ data were produced by the variation of the angle between the normal to the plane of the sample $n$ and magnetic field $H$ as a result of a change in the scalar product $(n, H)$, which in turn modulates the Hall signal by harmonic simple cosine law $\rho_H(\varphi) = \rho_{H0} + \rho_{H1}\cos\varphi$, where $\rho_{H0}$ is the constant bias term and $\rho_{H1}$ is the main component of the Hall signal. The amplitude of the harmonic term $\rho_{H1}$ was used to calculate the Hall coefficient $R_H = \rho_{H1}/H$. The uncertainty of the Hall measurements was estimated to be 5%. For all the specimens current of 1 or 10 μA and copper electrodes were used. The estimated uncertainties are $5\% \pm 0.5$ μV/K for the Seebeck coefficient, 2% for the electrical conductivity, and 8% for the thermal conductivity. It should be noted that in the course of this work the electrical and thermal transport properties were measured in different directions. According to our data and data reported by L. Jiang *et al*., this may lead to an additional limited misleading of the $zT$ of the order of $5 - 10$ % due to possible preferential grain orientation in BiCuSeO oxyselenides [49,58]. Thus, the total uncertainty for the figure of merit, $zT$, is approximately $20 - 25\%$.

## Results and discussion

X-ray diffraction analysis and Rietveld refinement confirmed the formation of the BiCuSeO phase that crystallizes in the ZrSiCuAs structure with the tetragonal *P4/nmm* space group for both SS and RS series of the samples (Fig. 1). Within the detection limit of the XRD technique, it seems that the RS samples are single phase (Fig. 1b), whereas the main peaks of $Bi_2O_3$ and $Cu_{2-x}Se$ are discernible at ~26.6° and ~28.3°, respectively, for SS samples (Fig. 1a). Nevertheless, $Bi_2O_3$ and $Cu_{2-x}Se$ account for less than 3 vol.% and 5 vol.%, respectively (see Table S2 and Fig. S1). According to the previously reported data, the increase of the secondary phase content with Sm content may originate from the Cu or/and Bi vacancies formed upon doping [59,60]. Thus, additional *p*-type doping can be expected for Sm-containing samples.

**Table 1.** Lattice parameters from Rietveld refinement of XRD data and elemental ratios obtained by EDS analysis for the $Bi_{1-x}Sm_xCuSeO$ ($x = 0 - 0.08$) samples prepared by SS and RS methods

| Nominal composition | SS route | | | | RS route | | | |
|---|---|---|---|---|---|---|---|---|
| | $a, b$ (Å) | $c$ (Å) | (Bi + Sm):Cu:Se | Sm | $a, b$ (Å) | $c$ (Å) | (Bi + Sm):Cu:Se | Sm |
| BiCuSeO [61] | 3.9213(1) | 8.9133(5) | n/a | n/a | n/a | n/a | n/a | n/a |
| BiCuSeO | 3.9221(1) | 8.9164(5) | 1.04:1.00:0.96 | 0.00 | 3.9263(2) | 8.9185(2) | 1.04:0.99:0.97 | 0.00 |
| $Bi_{0.98}Sm_{0.02}CuSeO$ | 3.9226(1) | 8.9194(3) | 1.03:0.99:0.98 | 0.02 | 3.9275(3) | 8.9211(4) | 1.02:0.98:1.00 | 0.02 |
| $Bi_{0.96}Sm_{0.04}CuSeO$ | 3.9253(5) | 8.9259(1) | 1.03:1.00:0.96 | 0.03 | 3.9293(2) | 8.9273(2) | 1.02:0.98:1.00 | 0.03 |
| $Bi_{0.94}Sm_{0.06}CuSeO$ | 3.9289(6) | 8.9303(7) | 1.03:0.99:0.99 | 0.05 | 3.9311(5) | 8.9316(4) | 1.02:0.98:1.00 | 0.05 |
| $Bi_{0.94}Sm_{0.08}CuSeO$ | 3.9290(7) | 8.9344(5) | 1.06:0.98:0.96 | 0.07 | 3.9313(2) | 8.9350(6) | 1.02:0.99:0.99 | 0.07 |
| SmCuSeO [5] | 3.9563(1) | 8.7164(1) | n/a | n/a | n/a | n/a | n/a | n/a |

The refined cell parameters of BiCuSeO are in good agreement with previously reported values (see Table 1) [61]. Without regard to synthesis technique, lattice parameters, *a* and *c*, for both SS and RS series increase with increasing Sm content. On the one hand, the ionic radius of $Bi^{3+}$ (1.03 Å) is quite close but still larger than those for $Sm^{3+}$ (0.96 Å) [62]. Thus, intuitively it seems that the decrease of the lattice parameters should be expected upon Bi substitution according to Vegard's law. On the other hand, one also should consider that $Sm^{2+}$ ions can be present in the system with a much larger ionic radius of 1.22 Å causing an increase of *a* and *c* [33,62]. Moreover, according to the crystallographic data collected for *M*CuSeO (*M* = La, Bi, Ce, Nd, Gd, Tb, Dy, Ho, and Y) the lattice parameter *a* should slightly increase (for less than 1%) even for $Sm^{3+}$ to $Bi^{3+}$ substitution as shown in Fig. S2 [5,63–68]. The presence of vacancies may also contribute to the lattice parameters evolution [25,69]. Thus, while there is a clear correlation between the increase of the lattice parameters and actual Sm content (see Table 1 and Fig. S3), it is difficult to unambiguously attribute the lattice parameters changes to some particular reason. It was also revealed that for some samples the actual Sm content is slightly lower than the nominal one (Table 1). Moreover, a negligible amount of $Sm_2O_3$ precipitates were detected by EDS analysis in RS samples with $x \geqslant 0.04$. For the SS samples traces of $Bi_2O_3$ (for the samples with $x \geqslant 0.02$) and $Cu_{2-x}Se$ (for the samples with $x \geqslant 0.04$) were also detected, which is consistent with XRD analysis (see Table S2).

Finally, all the samples exhibited fine and dense microstructure with randomly arranged lath-like grains which are typical for oxychalcogenides (see Fig. 2). According to SEM micrographs of the fractured cross-sections an average grain size (platelet thickness) is around 300 – 700 nm for SS and 400 – 1000 nm for RS samples, respectively (Table S3). Such difference can be attributed to the several intermediate ball milling steps used for the preparation of the SS series, while short and low-energy ball milling step was used during the preparation of the RS series [70]. It should be noted that in semiconductors, the microstructure-property relationship is quite strong. In particular, the grain size directly affects the degree of scattering of charge carriers ($\mu_H \propto L_g$, where $\mu_H$ is the Hall charge carrier mobility and $L_g$ is the grain size [71,72]) and acoustic phonons ($\kappa_{lat} \propto L_g$, where $\kappa_{lat}$ is the lattice thermal conductivity [73,74]), and thus the electrical and thermal properties of the material. Porosity, in this regard, also plays a crucial role. According to the effective medium model, even a small volume fraction of pores will increase both the thermal and electrical resistance of the material [75]. In addition, the usage of different preparation techniques may introduce new nano- or microstructures in the material, such as nano-pores, nanoparticles, amorphous regions, or different degree of preferential grains orientation. Thus, considering all the above mentioned, absolute values of the transport properties as well as its temperature dependence behaviour can be significantly tuned by using different material synthesis or processing techniques mainly due to the features of the formed nano- or microstructure (see, for example, Fig. 8 in Ref. [76], where the transport properties of pristine BiCuSeO obtained by different techniques are shown).

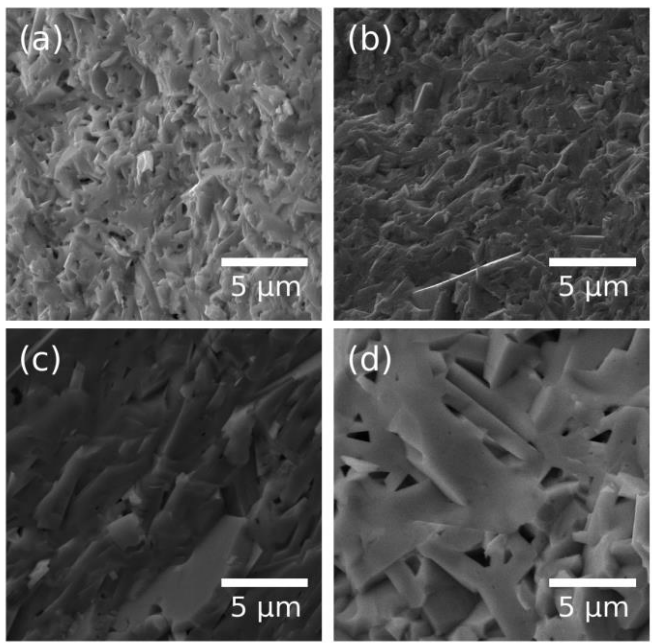

**Figure 2.** Scanning electron micrographs of a fractured surface of (a, c) BiCuSeO and (b, d) $Bi_{0.96}Sm_{0.04}CuSeO$ prepared by (a, b) SS and (c, d) RS method.

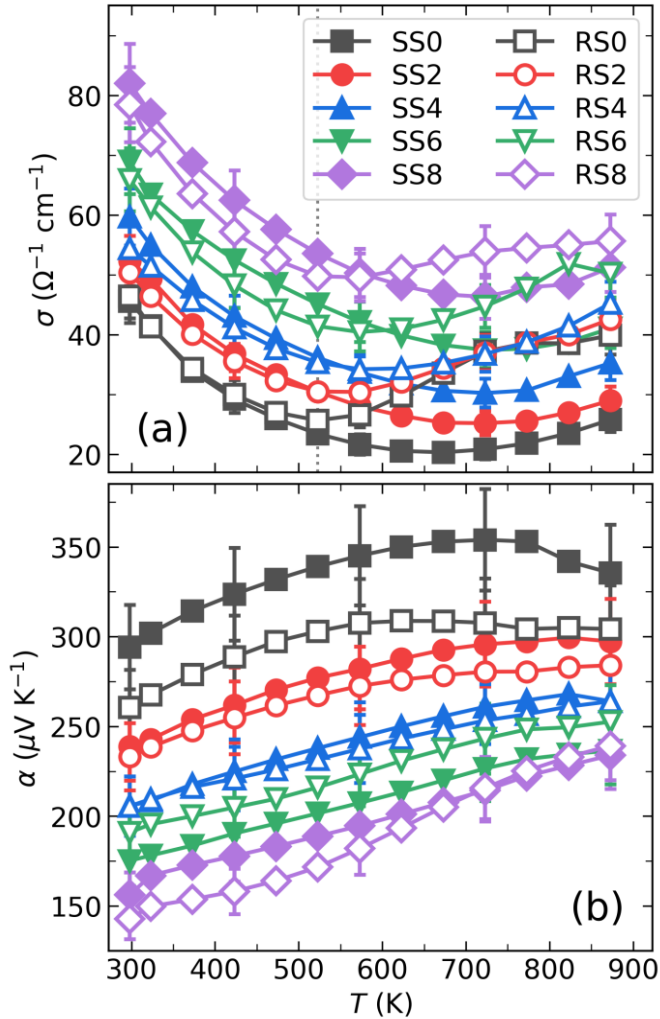

**Figure 3.** Temperature dependence of (a) the electrical conductivity, $\sigma$, (b) the Seebeck coefficient, $\alpha$, for the Bi$_{1-x}$Sm$_x$CuSeO ($x = 0 - 0.08$) samples prepared by SS and RS methods (solid and open symbols, respectively).

**Table 2.** Room-temperature Hall carrier concentration, Hall carrier mobility and the density of states effective mass calculated within the effective mass model for Bi$_{1-x}$Sm$_x$CuSeO ($x = 0, 0.04, 0.08$) samples prepared by SS and RS methods

| Code | $p_H$ (cm$^{-3}$) | $\mu_H$ (cm$^2$ V$^{-1}$ s$^{-1}$) | $m^*$ ($m_e$) |
|---|---|---|---|
| SS0 | 8.2·10$^{19}$ | 3.41 | 2.78 |
| SS4 | 1.6·10$^{20}$ | 2.38 | 2.93 |
| SS8 | 2.3·10$^{20}$ | 2.23 | 3.03 |
| RS0 | 7.8·10$^{19}$ | 3.69 | 2.38 |
| RS4 | 1.2·10$^{20}$ | 2.80 | 2.50 |
| RS8 | 2.4·10$^{20}$ | 2.07 | 2.76 |

Looking closer at the XRD pattern intensities of SS and RS bulks (Fig. 1) reveals that there is a noticeable difference in them. While there is no significant difference in intensities for XRD patterns of powders, for bulk samples the difference in intensities is noticeable and, first of all, it can be attributed to a different degree of preferential grains orientation (Fig. S4). This also agrees well with the calculated relative pole density values for the Bi$_{0.94}$Sm$_{0.06}$CuSeO as an example. So, for the sample obtained by the SS route, the pole densities were $\Phi_{001} = 0.55$ and $\Phi_{hk0} = 1.21$, respectively, while for the RS sample $\Phi_{001} = 2.25$ and $\Phi_{hk0} = 0.58$. In other words, it can be stated with confidence that in the samples of the two series, the crystallites are oriented differently relative to the non-textured standard sample, which, in principle, should affect the electrical and thermal transport in the specimens as shown and discussed below (Fig. 3).

The electrical conductivity of all the samples decreases upon heating, indicating a degenerate behavior up to some temperature $T_i$ (depending on the sample) as shown in Fig. 3a. The electrical conductivity of the RS samples is similar to those of SS samples at low temperatures. However, an upturn temperature, $T_i$, seems to be much lower for the RS samples resulting in a slightly higher $\sigma$ at high temperatures. In general, for both series the electrical conductivity is increased by the Sm for Bi substitution and reached maximum values of ~80 $\Omega^{-1}$ cm$^{-1}$ at 300 K and ~53 $\Omega^{-1}$ cm$^{-1}$ at 873 K for Bi$_{0.92}$Sm$_{0.08}$CuSeO samples.

The Seebeck coefficient is positive in the investigated temperature range for all sample compositions exhibiting a *p*-type transport behavior (Fig. 3b). The pristine samples show a large Seebeck coefficient of more than 250 µV K$^{-1}$ at 300 K and even higher values of 300 – 340 µV K$^{-1}$ at 823 K. With the increase of Sm content, the temperature dependence of the Seebeck coefficient becomes more linear as expected for degenerate semiconductors. All the doped samples obtained by the RS technique exhibit similar values of $\alpha$ compared to those of SS samples in the whole temperature range. The value reaches ~240 µV K$^{-1}$ at 823 K for the highest samarium content. It should be mentioned that the electrical transport properties of all the samples were measured for several runs in order to ensure that the $\sigma$ and $\alpha$ values are reproducible. Moreover, the same measurements were performed for SS and RS Bi$_{0.94}$Sm$_{0.06}$CuSeO specimens after annealing for 24 h at 873 K to show that the samples are stable and are in their thermodynamic equilibrium state. All the values of $\sigma$ and $\alpha$ after annealing are in good agreement (in the range of uncertainty) with those before annealing (see Fig. S5).

According to previous reports, there are several possible reasons for such changes observed in the electrical transport properties of BiCuSeO doped with variable-valence Sm: (i) partial substitution of Bi$^{3+}$ by Sm$^{2+}$ [33], which will introduce more charge carriers ($h^\bullet$, holes) to the system according to the equation: $Bi^{\times}_{Bi} \to Sm'_{Cu} + h^\bullet$ [32]; (ii) the band structure tuning as can be expected for rare-earth doped oxychalcogenides [24]; (iii) the formation of crystal defects, such as anti-sites, vacancies, etc., also resulting in an enhancement of $p_H$ [76]. Unfortunately, within the framework of this work, we believe that it is impossible to unambiguously determine the ratio of $Sm^{2+}/Sm^{3+}$ ions, as was erroneously done in a previous report [33]. In general, the XPS spectrum for the $Bi_{0.92}Sm_{0.08}CuSeO$ sample (Fig. S6) is in good agreement with the results presented in previous reports (for details, see ESI). However, due to a small difference in the binding energy between $Sm^{2+}$ and $Sm^{3+}$, it is impossible to perform a qualitative deconvolution of the peaks without a good $Sm^{2+}$ standard. Nevertheless, according to the results of Hall measurements (Fig. S7), the concentration of charge carriers increases with the increase in the Sm content (Table 2), which is generally consistent with previously reported data (Table S4) [33,34]. For both the SS and RS series of samples, the pattern of change in transport properties upon Sm content is similar and agrees with those previously reported for Sm-doped BiCuSeO oxyselenides (Fig. S8) [33,34]. However, two aspects should be mentioned when our results are compared with previously reported ones. Firstly, despite a similar trend in the change of the electrical conductivity upon Sm doping (Fig. S8a), the mechanism that causes this change is reported to be different in previous reports. While our data and data from previous reports agree that charge carrier concentration increases along with the increase of Sm doping level, the carrier mobility data is controversial. On the one hand, in the paper by Feng *et al*. [34], the mobility of the undoped BiCuSeO sample was ~7 $cm^2\ V^{-1}\ s^{-1}$ and increased significantly with an increase in the Sm content. On the other hand, in the paper by Kang *et al*. [33], the mobility for all the samples was in the range of $2 - 4\ cm^2\ V^{-1}\ s^{-1}$ and was only slightly increased with Sm doping. For a reasonable conclusion on the origin of such differences, it is necessary to have more information about the samples, their preparation, and microstructure. The experimental data on Hall mobility obtained in this work (Table 2) are consistent with the results obtained by Kang *et al*. [33]. Secondly, the temperature dependences of the electrical conductivity of the samples presented in the literature also differ from what was observed in this work for Sm-doped BiCuSeO (Fig. S9). We think that this is most likely attributed to the features of the microstructure of the samples obtained by different recipes, as discussed earlier (Fig. S10 or Fig. 8 in Ref. [76]).

The total thermal conductivity, $\kappa$, of all the samples is close to $0.8 - 1.1\ W\ m^{-1}K^{-1}$ at room temperature and falls down to $0.5 - 0.8\ W\ m^{-1}K^{-1}$ at 873 K (Fig. 4a). The total thermal conductivity is the sum of the lattice thermal conductivity $\kappa_{lat}$ and the electronic thermal conductivity $\kappa_{el}$: $\kappa = \kappa_{lat} + \kappa_{el}$. The electronic thermal conductivity is proportional to the electrical conductivity according to the Wiedmann-Franz law $\kappa_{el} = \sigma LT$. Here, the Lorenz number, $L$, was estimated as a function of temperature from the experimental values of the Seebeck coefficient in the framework of the effective mass model (Fig. S11a). Both $\kappa_{lat}$ and $\kappa_{el}$ are increasing upon doping (Fig. S11b, S12b) leading to the rise of $\kappa$. Firstly, increased $\sigma$ for Sm doped samples results in raise of $\kappa_{el}$. Secondly, for materials with a complex primitive cell at temperatures above Debye temperature ($\theta_D$ = 243 K for BiCuSeO) [77] lattice thermal conductivity can be roughly modeled as $\kappa_{lat} \propto A/T + B$, with contributions from Umklapp scattering, $A \propto \dfrac{\bar{M}v_a^3}{\gamma^2 n_{at}^{1/3} V_{at}^{2/3}}$, and boundary scattering, $B \propto \dfrac{v_a L_g}{V_{at} n_{at}}$, where $\bar{M}$ is the average mass, $v_a$ is the average sound velocity, $\gamma$ is the Grüneisen parameter, $n_{at}$ is the number of atoms per primitive cell, $V_{at}$ is the volume per atom, $L_g$ is the grain size [73]. Thus, considering the mass difference between Sm (150.36 g $mol^{-1}$) and Bi (208.98 g $mol^{-1}$) atoms and higher sound velocity in SmCuSeO [78] (see Table S5), the lattice thermal conductivity should increase upon substitution of Bi to Sm, which is in line with previous reports [33,34]. Moreover, according to this model, slightly higher thermal conductivity for the RS series in comparison with the SS

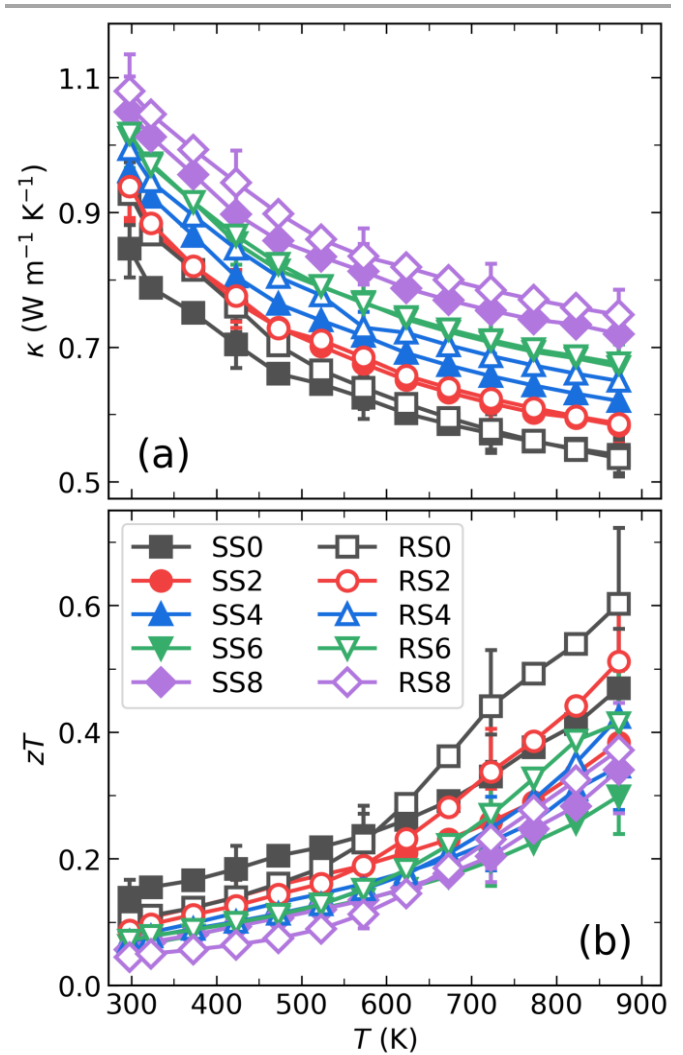

**Figure 4.** Temperature dependence of (a) the total thermal conductivity and (b) the figure of merit $zT$ for the $Bi_{1-x}Sm_xCuSeO$ ($x$ = 0 – 0.08) samples prepared by SS and RS methods (solid and open symbols, respectively).

series can be attributed to a smaller average grain size of the latter.

The combination of the electrical and thermal transport properties allows one to estimate the figure of merit, $zT$, as a function of temperature for the $Bi_{1-x}Sm_xCuSeO$ samples (Fig. 4b). The pristine samples exhibit values in a range of 0.5 – 0.6 at 873 K, which is in good agreement with many previous reports [77]. However, the $zT$ values of the doped samples were lower than that of pristine BiCuSeO for both SS and RS series. This can be explained by considering the so-called quality factor, $\beta \propto \mu_w T^{5/2}/\kappa_{lat}$ (here $\mu_w$ is the weighted mobility related to the drift mobility as $\mu_w \approx \mu(m_d^*/m_e)^{3/2}$, with $m_d^*$ is the density of states effective mass and $m_e$ is the electron mass), which determines the maximum $zT$ value achievable for a given material at a given temperature [79]. On the one hand, the effective mass increases with doping due to the increase in the ionicity of the $[Bi_{2-2x}Sm_xO_2]^{2(1-x)+}$ layers [80,81] (see Table 2 and Table S6), which is beneficial for the weighted mobility. On the other hand, assuming a single parabolic band and deformation potential scattering $\mu \propto T^{3/2}(m_d^*)^{-5/2}$ and thus the increase in the density of states effective mass would actually result in $\mu_w$ reduction with doping (Fig. S8g). Thus, considering the increased lattice thermal conductivity, which is not compensated by the weighted mobility, the quality factor is reduced upon doping as well as $zT$ (Fig. S8).

## Conclusions

The thermoelectric properties of the $Bi_{1-x}Sm_xCuSeO$ ($0 \leq x \leq 0.08$) samples prepared by two-step reactive sintering and solid-state reaction route have been studied. The Sm for Bi substitution increases the electrical conductivity while moderately decreasing the Seebeck coefficient. The intrinsically low thermal conductivity of BiCuSeO related to large anharmonicity, weak interatomic bonding between the layers, and large atomic weight of constituents [77], is noticeably increased by the Sm doping mainly due to the lower weight of Sm. Finally, with the combination of deteriorated weighted mobility and increased lattice thermal conductivity the figure of merit $zT$ is decreasing upon Sm for Bi substitution. Despite the obtained low $zT$ value, it is instructive to clarify these rare earth doping effects, and the reactive sintering approach developed to synthesize undoped and doped BiCuSeO oxyselenides is in practice an interesting playground, which can be extended to other oxychalcogenides in the future.

## CRediT authorship contribution statement

**Andrei Novitskii:** Conceptualization, Methodology, Formal analysis, Investigation, Writing – original draft, Writing – Review & Editing, Visualization. **Illia Serhiienko:** Conceptualization, Methodology, Investigation. **Sergey Novikov:** Investigation. **Kirill Kuskov:** Investigation. **Daria Pankratova:** Investigation. **Tatyana Sviridova:** Investigation, Formal analysis. **Andrei Voronin:** Resources, Project administration, Funding acquisition. **Aleksei Bogach:** Investigation. **Elena Skryleva:** Investigation **Yuriy Parkhomenko:** Resources. **Alexander Burkov:** Resources, Data curation, Writing – Review & Editing. **Takao Mori:** Resources, Writing – Review & Editing. **Vladimir Khovaylo:** Resources, Writing – Review & Editing, Supervision.

## Conflicts of interest

The authors declare no competing financial interest.

## Acknowledgements

The study was carried out with financial support from the Russian Science Foundation, project No. 19-79-10282. T.M. acknowledges JSPS KAKENHI JP16H06441 and JST Mirai JPMJMI19A1.

## Data availability

The data that support the findings of this study are available from the corresponding author upon reasonable request.

## References

[1] V. Johnson, W. Jeitschko, J. Solid State Chem. 11 (1974) 161–166.

[2] R. Pöttgen, D. Johrendt, Zeitschrift Fur Naturforsch. - Sect. B J. Chem. Sci. 63 (2008) 1135–1148.

[3] V. V. Bannikov, A.L. Ivanovskii, J. Struct. Chem. 56 (2015) 148–162.

[4] M. Palazzi, C. Carcaly, J. Flahaut, J. Solid State Chem. 35 (1980) 150–155.

[5] T. Ohtani, M. Hirose, T. Sato, K. Nagaoka, M. Iwabe, Jpn. J. Appl. Phys. 32 (1993) 316–318.

[6] Y. Takano, K. Yahagi, K. Sekizawa, Phys. B Condens. Matter 206–207 (1995) 764–766.

[7] K. Ueda, S. Inoue, S. Hirose, H. Kawazoe, H. Hosono, Appl. Phys. Lett. 77 (2000) 2701–2703.

[8] T. Ohtani, Y. Tachibana, Y. Fujii, J. Alloys Compd. 262–263 (1997) 175–179.

[9] L.D. Zhao, D. Berardan, Y.L. Pei, C. Byl, L. Pinsard-Gaudart, N. Dragoe, Appl. Phys. Lett. 97 (2010) 092118.

[10] P. Vaqueiro, G. Guélou, M. Stec, E. Guilmeau, A. V. Powell, J. Mater. Chem. A 1 (2013) 520–523.

[11] Y.-L. Pei, H. Wu, D. Wu, F. Zheng, J. He, J. Am. Chem. Soc. 136 (2014) 13902–13908.

[12] J. Sui, J. Li, J. He, Y.-L. Pei, D. Berardan, H. Wu, N. Dragoe, W. Cai, L.-D. Zhao, Energy Environ. Sci. 6 (2013) 2916.

[13] Y. Liu, L.-D. Zhao, Y. Zhu, Y. Liu, F. Li, M. Yu, D.-B. Liu, W. Xu, Y.-H. Lin, C.-W. Nan, Adv. Energy Mater. 6 (2016) 1502423.

[14] G.-K. Ren, J.-L. Lan, L.-D. Zhao, C. Liu, H. Yuan, Y. Shi, Z. Zhou, Y.-H. Lin, Mater. Today 29 (2019) 68–85.

[15] A.F. Ioffe, Semiconductor Thermoelements, and Thermoelectric Cooling, Infosearch, London, 1957.

[16] J. He, T.M. Tritt, Science (80-. ). 357 (2017) eaak9997.
[17] T. Mori, Small 13 (2017) 1702013.
[18] J. Mao, Z. Liu, J. Zhou, H. Zhu, Q. Zhang, G. Chen, Z. Ren, Adv. Phys. 67 (2018) 69–147.
[19] V.I. Fistul', Heavily Doped Semiconductors, Springer New York, Boston, MA, 1995.
[20] J. Li, J. Sui, Y. Pei, X. Meng, D. Berardan, N. Dragoe, W. Cai, L.-D. Zhao, J. Mater. Chem. A 2 (2014) 4903–4906.
[21] J.-L. Lan, Y.-C. Liu, B. Zhan, Y.-H. Lin, B. Zhang, X. Yuan, W. Zhang, W. Xu, C.-W. Nan, Adv. Mater. 25 (2013) 5086–5090.
[22] F. Li, T.-R. Wei, F. Kang, J.-F. Li, J. Mater. Chem. A 1 (2013) 11942–11949.
[23] Y. Liu, J. Lan, W. Xu, Y. Liu, Y.L. Pei, B. Cheng, D.B. Liu, Y.H. Lin, L.D. Zhao, Chem. Commun. 49 (2013) 8075–8077.
[24] Y. Liu, J. Ding, B. Xu, J. Lan, Y. Zheng, B. Zhan, B. Zhang, Y. Lin, C. Nan, Appl. Phys. Lett. 106 (2015) 233903.
[25] Y. Liu, L.-D. Zhao, Y. Liu, J. Lan, W. Xu, F. Li, B.-P. Zhang, D. Berardan, N. Dragoe, Y.-H. Lin, C.-W. Nan, J.-F. Li, H. Zhu, J. Am. Chem. Soc. 133 (2011) 20112–20115.
[26] Z. Li, C. Xiao, S. Fan, Y. Deng, W. Zhang, B. Ye, Y. Xie, J. Am. Chem. Soc. 137 (2015) 6587–6593.
[27] F. Li, M. Ruan, Y. Chen, W. Wang, J. Luo, Z. Zheng, P. Fan, Inorg. Chem. Front. 6 (2019) 799–807.
[28] Y. Chen, K.-D. Shi, F. Li, X. Xu, Z.-H. Ge, J. He, J. Am. Ceram. Soc. 102 (2019) 5989–5996.
[29] L. Pan, Y. Lang, L. Zhao, D. Berardan, E. Amzallag, C. Xu, Y. Gu, C. Chen, L.-D. Zhao, X. Shen, Y. Lyu, C. Lu, Y. Wang, J. Mater. Chem. A 6 (2018) 13340–13349.
[30] Z. Yin, Z. Liu, Y. Yu, C. Zhang, P. Chen, J. Zhao, P. He, X. Guo, ACS Appl. Mater. Interfaces (2021) acsami.1c19266.
[31] Y. Gu, X. Shi, L. Pan, W. Liu, Q. Sun, X. Tang, L. Kou, Q. Liu, Y. Wang, Z. Chen, Adv. Funct. Mater. (2021) 2101289.
[32] H. Kang, J. Li, Y. Liu, E. Guo, Z. Chen, D. Liu, G. Fan, Y. Zhang, X. Jiang, T. Wang, J. Mater. Chem. C 6 (2018) 8479–8487.
[33] H. Kang, X. Zhang, Y. Wang, J. Li, D. Liu, Z. Chen, E. Guo, X. Jiang, T. Wang, Mater. Res. Bull. 126 (2020) 110841.
[34] B. Feng, X. Jiang, Z. Pan, L. Hu, X. Hu, P. Liu, Y. Ren, G. Li, Y. Li, X. Fan, Mater. Des. 185 (2020) 108263.
[35] X. Zhou, Y. Yan, X. Lu, H. Zhu, X. Han, G. Chen, Z. Ren, Mater. Today 21 (2018) 974–988.
[36] S. Il Kim, K.H. Lee, H.A. Mun, H.S. Kim, S.W. Hwang, J.W. Roh, D.J. Yang, W.H. Shin, X.S. Li, Y.H. Lee, G.J. Snyder, S.W. Kim, Science (80-. ). 348 (2015) 109–114.
[37] X.L. Su, F. Fu, Y.G. Yan, G. Zheng, T. Liang, Q.Q.J. Zhang, X. Cheng, D.W. Yang, H. Chi, X.F. Tang, Q.Q.J. Zhang, C. Uher, Nat. Commun. 5 (2014) 1–7.
[38] Y. Lan, A.J. Minnich, G. Chen, Z. Ren, Adv. Funct. Mater. 20 (2010) 357–376.
[39] G.-K. Ren, J.-L. Lan, S. Butt, K.J. Ventura, Y.-H. Lin, C.-W. Nan, RSC Adv. 5 (2015) 69878–69885.
[40] V. Pele, C. Barreteau, D. Berardan, L. Zhao, N. Dragoe, J. Solid State Chem. 203 (2013) 187–191.
[41] G.-K. Ren, S.-Y. Wang, Y.-C. Zhu, K.J. Ventura, X. Tan, W. Xu, Y.-H. Lin, J. Yang, C.-W. Nan, Energy Environ. Sci. 10 (2017) 1590–1599.
[42] O. Guillon, J. Gonzalez-Julian, B. Dargatz, T. Kessel, G. Schierning, J. Räthel, M. Herrmann, Adv. Eng. Mater. 16 (2014) 830–849.
[43] A.S. Mukasyan, A.S. Rogachev, D.O. Moskovskikh, Z.S. Yermekova, Ceram. Int. 48 (2022) 2988–2998.
[44] D.Y. Nhi Truong, H. Kleinke, F. Gascoin, Intermetallics 66 (2015) 127–132.
[45] E. Yaprintseva, A. Vasil'ev, M. Yaprintsev, O. Ivanov, Mater. Lett. 309 (2022) 131416.
[46] N.K. Upadhyay, N.S. Chauhan, L.A. Kumaraswamidhas, K.K. Johari, B. Gahtori, S. Bathula, R. Reddy, Y. V. Kolen'ko, S.R. Dhakate, A. Dhar, Mater. Lett. 265 (2020) 127428.
[47] S. Choudhary, S. Muthiah, S.R. Dhakate, ACS Appl. Energy Mater. 5 (2022) 549–556.
[48] A. Novitskii, G. Guélou, A. Voronin, T. Mori, V. Khovaylo, Scr. Mater. 187 (2020) 317–322.
[49] A. Novitskii, I. Serhiienko, S. Novikov, Y. Ashim, M. Zheleznyi, K. Kuskov, D. Pankratova, P. Konstantinov, A. Voronin, O. Tretiakov, T. Inerbaev, A. Burkov, V. Khovaylo, (2021) http://arxiv.org/abs/2104.10509.
[50] E. V. Shelekhov, T.A. Sviridova, Met. Sci. Heat Treat. 42 (2000) 309–313.
[51] Y.S. Umanskii, Y.A. Skakov, A.N. Ivanov, L.N. Rastorguev, Crystallography, X-Ray Analysis and Electron Microscopy (in Russian), Metallurgy, Moscow, 1982.
[52] R.E. Dinnebier, S.J.L. Billinge, eds., Powder Diffraction, Royal Society of Chemistry, Cambridge, 2008.
[53] J.A. Kaduk, S.J.L. Billinge, R.E. Dinnebier, N. Henderson, I. Madsen, R. Černý, M. Leoni, L. Lutterotti, S. Thakral, D. Chateigner, Nat. Rev. Methods Prim. 1 (2021) 77.
[54] J.I. Goldstein, D.E. Newbury, J.R. Michael, N.W.M. Ritchie, J.H.J. Scott, D.C. Joy, Scanning Electron Microscopy and X-Ray Microanalysis, Springer New York, New York, NY, 2018.
[55] H. Nakajima, (2021).
[56] A.T. Burkov, A. Heinrich, P.P. Konstantinov, T. Nakama, K. Yagasaki, Meas. Sci. Technol. 12 (2001) 264–272.
[57] C. Kittel, ed., Introduction to Solid State Physics, 8th ed., John Wiley & Sons, Inc, 2005.
[58] L. Jiang, L. Han, C. Lu, S. Yang, Y. Liu, H. Jiang, Y. Yan, X. Tang, D. Yang, ACS Appl. Mater. Interfaces 13 (2021) 11977–11984.
[59] M. Ishizawa, Y. Yasuzato, H. Fujishiro, T. Naito, H. Katsui, T. Goto, J. Appl. Phys. 123 (2018) 245104.

[60] S. Das, A. Ramakrishnan, K. Chen, J. Phys. D. Appl. Phys. 51 (2018) 035501.
[61] L.N. Kholodkovskaya, L.G. Akselrud, A.M. Kusainova, V.A. Dolgikh, B.A. Popovkin, Mater. Sci. Forum 133–136 (1993) 693–696.
[62] R.D. Shannon, Acta Crystallogr. Sect. A 32 (1976) 751–767.
[63] W.J. Zhu, Y.Z. Huang, C. Dong, Z.X. Zhao, Mater. Res. Bull. 29 (1994) 143–147.
[64] A.M. Kusainova, P.S. Berdonosov, L.G. Akselrud, L.N. Kholodkovskaya, V.A. Dolgikh, B.A. Popovkin, J. Solid State Chem. 112 (1994) 189–191.
[65] P.S. Berdonosov, A.M. Kusainova, L.N. Kholodkovskaya, V.A. Dolgikh, L.G. Akselrud, B.A. Popovkin, J. Solid State Chem. 118 (1995) 74–77.
[66] Y. Ohki, S. Komatsuzaki, Y. Takahashi, K. Takase, Y. Takano, K. Sekizawa, in: AIP Conf. Proc., AIP, 2006, pp. 1309–1310.
[67] C. Barreteau, D. Bérardan, L. Zhao, N. Dragoe, J. Mater. Chem. A 1 (2013) 2921–2926.
[68] C.-L. Hsiao, X. Qi, Acta Mater. 102 (2016) 88–96.
[69] C. Barreteau, D. Bérardan, E. Amzallag, L. Zhao, N. Dragoe, Chem. Mater. 24 (2012) 3168–3178.
[70] S. Mizuno, M. Ishizawa, H. Fujishiro, T. Naito, H. Katsui, T. Goto, Jpn. J. Appl. Phys. 55 (2016) 115801.
[71] P.S. Kireev, Semiconductor Physics, 2nd ed., Mir, Moscow, 1978.
[72] M. Grundmann, The Physics of Semiconductors, Springer International Publishing, Cham, 2016.
[73] E.S. Toberer, A. Zevalkink, G.J. Snyder, J. Mater. Chem. 21 (2011) 15843.
[74] T.M. Tritt, ed., Thermal Conductivity: Theory, Properties, and Applications, Springer US, New York, 2004.
[75] Philos. Trans. R. Soc. London. Ser. A, Contain. Pap. a Math. or Phys. Character 203 (1904) 385–420.
[76] A.P. Novitskii, V. V. Khovaylo, T. Mori, Nanobiotechnology Reports 16 (2021) 294–307.
[77] L.-D. Zhao, J. He, D. Berardan, Y. Lin, J.-F. Li, C. Nan, N. Dragoe, Energy Environ. Sci. 7 (2014) 2900–2924.
[78] J. He, Y. Xia, W. Lin, K. Pal, Y. Zhu, M.G. Kanatzidis, C. Wolverton, (2021) arXiv:2107.04955.
[79] R.P. Chasmar, R. Stratton, J. Electron. Control 7 (1959) 52–72.
[80] J.C. Phillips, Rev. Mod. Phys. 42 (1970) 317–356.
[81] L.M. Minges, Electronic Materials Handbook: Packaging, Volume I, CRC Press, 1989.

# Thermoelectric properties of Sm-doped BiCuSeO oxyselenides fabricated by two-step reactive spark plasma sintering

Andrei Novitskii,[a,b,*] Illia Serhiienko,[a,c,d] Sergey Novikov,[b] Kirill Kuskov,[a] Daria Pankratova,[a,‡] Tatyana Sviridova,[a] Andrei Voronin,[a] Aleksei Bogach[e], Elena Skryleva,[a] Yuriy Parkhomenko,[a,f] Alexander Burkov,[b] Takao Mori,[c,d] Vladimir Khovaylo[a]

[a] National University of Science and Technology MISIS, 119049 Moscow, Russian Federation

[b] Ioffe Institute, 194021 St. Petersburg, Russian Federation

[*] E-mail: novitskiy@misis.ru

[c] National Institute for Materials Science (NIMS), International Center for Materials Nanoarchitectonics (WPI-MANA), 305-0044 Tsukuba, Japan

[d] University of Tsukuba, Graduate School of Pure and Applied Sciences, 305-8671 Tsukuba, Japan

[e] Prokhorov General Physics Institute of the Russian Academy of Sciences, 119991 Moscow, Russian Federation

[f] JSC "Giredmet", 111524 Moscow, Russian Federation

[‡] Present address: Luleå University of Technology, 97187 Luleå, Sweden

**Table S1.** Melting and boiling points, bulk density, electrical and thermal conductivities of the materials used for the Bi$_{1-x}$Sm$_x$CuSeO ($x = 0 - 0.08$) synthesis.[*]

| Material | Melting point (K) | Boiling point (K) | Density (g cm$^{-3}$) | Electrical resistivity (μΩ cm) | Thermal conductivity (W m$^{-1}$K$^{-1}$) |
|---|---|---|---|---|---|
| Bi | 544.4 | 1837 | 9.79 | 110 (273 K) | ~8 (273 K) [1] |
| Cu | 1357.8 | 2833 | 8.96 | 1.5 (273 K) | 399 (300 K) [1] |
| Se | 494 | 958 | 4.81 | 8·10$^9$ (753 K) [2] | 0.52 (300 K) [1] |
| Bi$_2$O$_3$ | 1098 | 2163 | 8.90 | 3·10$^{12}$ (300 K) [3] | ~0.9 (500 K) [4] |
| Sm$_2$O$_3$ | 2608 | 4053 | 7.60 | 5·10$^{13}$ (673 K) [5] | – |

[*]all the data are from 97$^{th}$ edition of CRC Handbook of Chemistry and Physics [6] unless otherwise stated

**Table S2.** Details of the phase composition of the Bi$_{1-x}$Sm$_x$CuSeO ($x = 0 - 0.08$) samples prepared by two-step solid-state reaction combined with ball milling and SPS (labeled as SS); by two-step reactive spark plasma sintering (samples were labeled as RS) according to the XRD pattern refinement.[*]

| Nominal composition | Phase composition (vol.%) | |
|---|---|---|
| | SS | RS |
| BiCuSeO | 100% BiCuSeO | 100% BiCuSeO |
| Bi$_{0.98}$Sm$_{0.02}$CuSeO | 99.1% BiCuSeO, 0.9% Bi$_2$O$_3$ | 99.4% BiCuSeO, 0.6% Bi$_2$O$_3$ |
| Bi$_{0.96}$Sm$_{0.04}$CuSeO | 98.8% BiCuSeO, 1.2% Bi$_2$O$_3$ | 100% BiCuSeO |
| Bi$_{0.94}$Sm$_{0.06}$CuSeO | 95.3% BiCuSeO, 2.6% Bi$_2$O$_3$, 2.1% Cu$_{2-x}$Se | 100% BiCuSeO |
| Bi$_{0.94}$Sm$_{0.08}$CuSeO | 93.1% BiCuSeO, 1.9% Bi$_2$O$_3$, 5.0% Cu$_{2-x}$Se | 100% BiCuSeO |

[*]The final refinement of XRD patterns was carried out assuming a tetragonal crystal system with a space group of *P4/nmm* (No. 129) and taking the pseudo-Voigt function for the peak profiles in the 2$\theta$ range of 10 – 110° (Cu$K_\alpha$ radiation with $\lambda$ = 1.54178 Å).

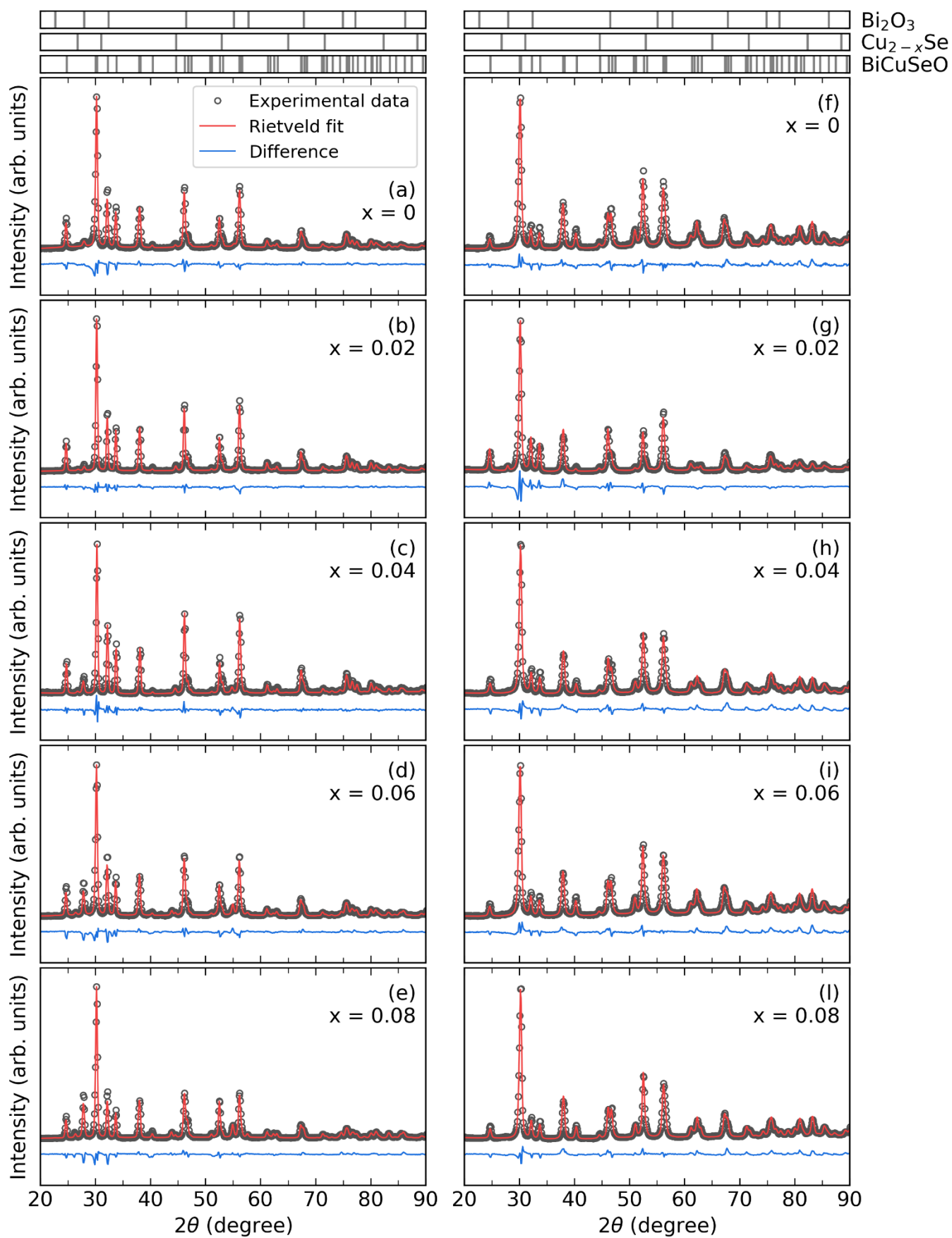

**Fig. S1.** Rietveld refinements of Bi$_{1-x}$Sm$_x$CuSeO ($x$ = 0 – 0.08) specimens obtained by SS (a – e) and RS (f – l) routes, respectively.

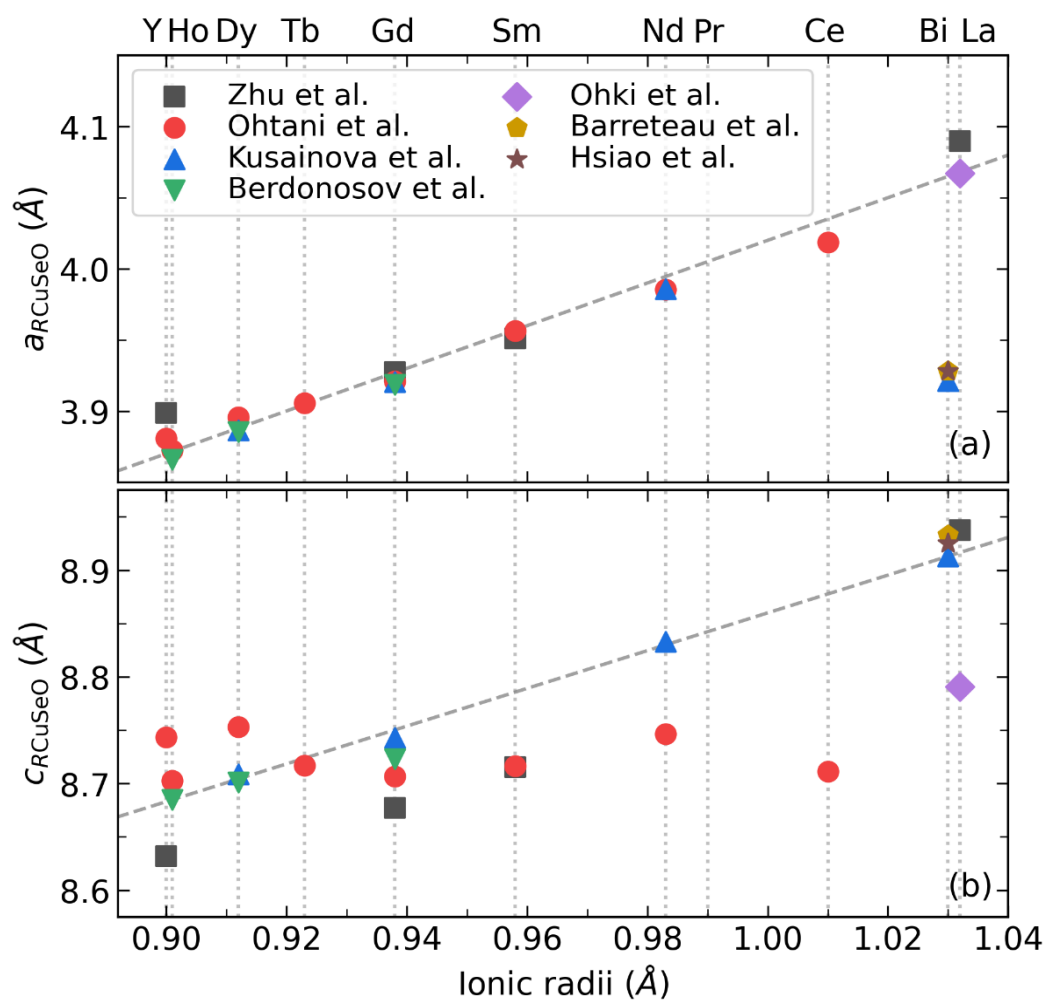

**Figure S2.** Dependence of lattice parameters (a) *a* and (b) *c* on ionic radii of *R* in *R*CuSeO (where *R* = Bi, La, Ce, Pr, Nd, Sm, Gd, Tb, Dy, Ho, Y). Data was collected from previous reports [7–13]. The dashed grey lines are guide to the eye.

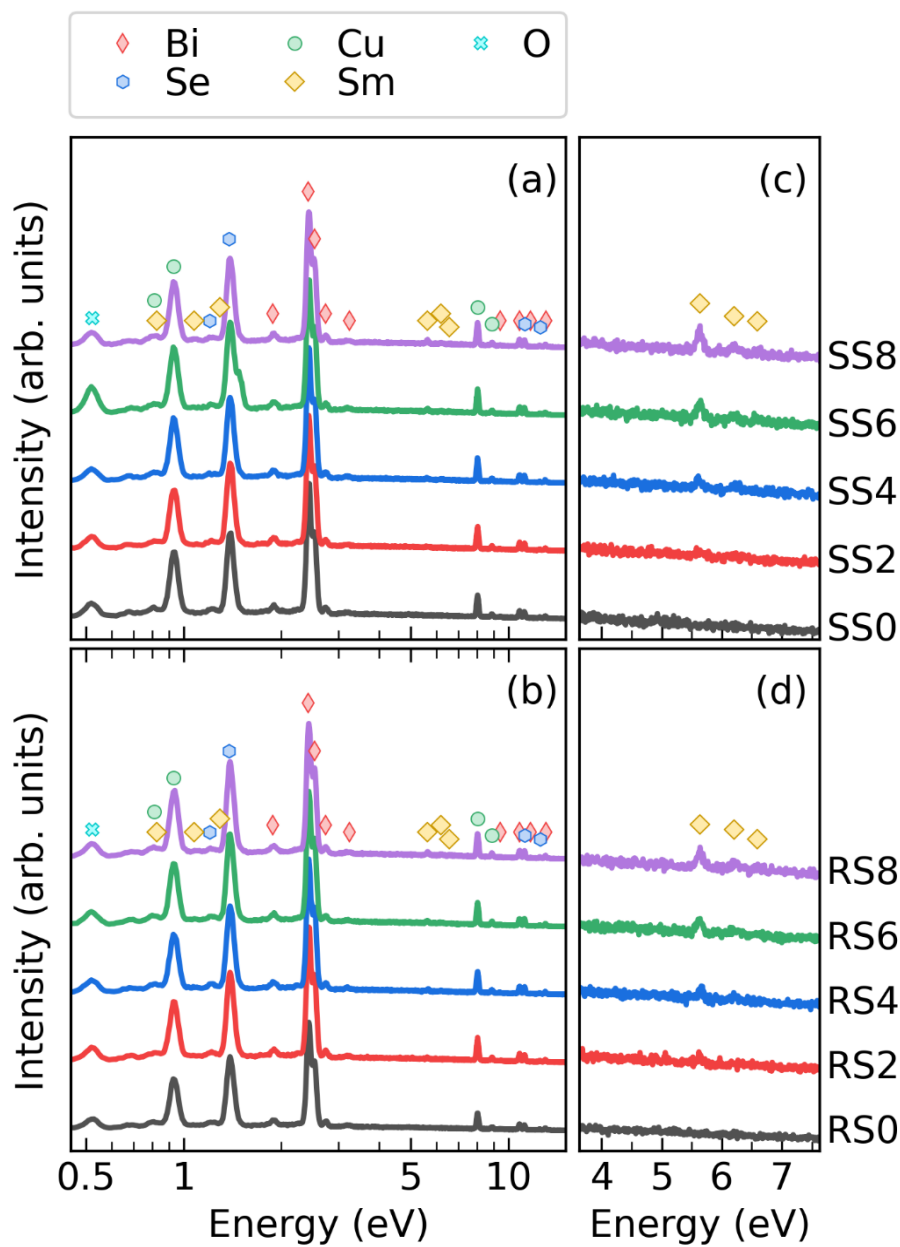

**Figure S3.** EDS patterns for Bi$_{1-x}$Sm$_x$CuSeO ($x = 0 - 0.08$) prepared by both (a) SS and (b) RS techniques. (c, d) Enlarged section of (a) and (b) with peaks of Sm corresponding to $L_{\alpha 1}$ (5.636 eV), $L_{\beta 1}$ (6.205 eV) and $L_{\beta 2.15}$ (6.587 eV) lines.

**Table S3.** Average grain size, $s_g$, estimated from scanning electron micrographs and bulk density, $d$, of the $Bi_{1-x}Sm_xCuSeO$ ($x = 0 - 0.08$) samples prepared by SS and RS routes.[*]

| Nominal composition | SS | | RS | |
|---|---|---|---|---|
| | $s_g$ (nm) | $d \pm 1\%$ (%) | $s_g$ (nm) | $d \pm 1\%$ (%) |
| BiCuSeO | 520 ± 50 | 94.2 | 720 ± 70 | 94.1 |
| $Bi_{0.98}Sm_{0.02}CuSeO$ | 500 ± 70 | 94.2 | 730 ± 90 | 94.0 |
| $Bi_{0.96}Sm_{0.04}CuSeO$ | 540 ± 100 | 93.9 | 770 ± 110 | 93.6 |
| $Bi_{0.94}Sm_{0.06}CuSeO$ | 520 ± 80 | 93.7 | 760 ± 100 | 93.7 |
| $Bi_{0.94}Sm_{0.08}CuSeO$ | 490 ± 90 | 93.4 | 710 ± 100 | 93.4 |

[*]Theoretical density of BiCuSeO estimated as $d_{th} = \sum_i n_i A_i/(VN_A) = 8.910$ g cm$^{-3}$, where $n_i$ is the number of each atom (Bi, Cu, Se and O) in the unit cell, $n_i = 2$, $A_i$ is the atomic weight of each atom in the unit cell, $V$ is the volume of the unit cell, $N_A$ is the Avogadro's number ($6.022 \cdot 10^{23}$ atoms/mol).

**Figure S4.** X-ray diffraction patterns for $Bi_{0.94}Sm_{0.06}CuSeO$ in the form of (a) powder and (b) bulk, respectively. The Bragg's reflections for BiCuSeO phase are indicated by grey ticks on the top part of the figure. $Bi_2O_3$ and $Cu_{2-x}Se$ secondary phases are indicated by black solid triangle (▼) and thin diamond (♦) symbols, respectively.

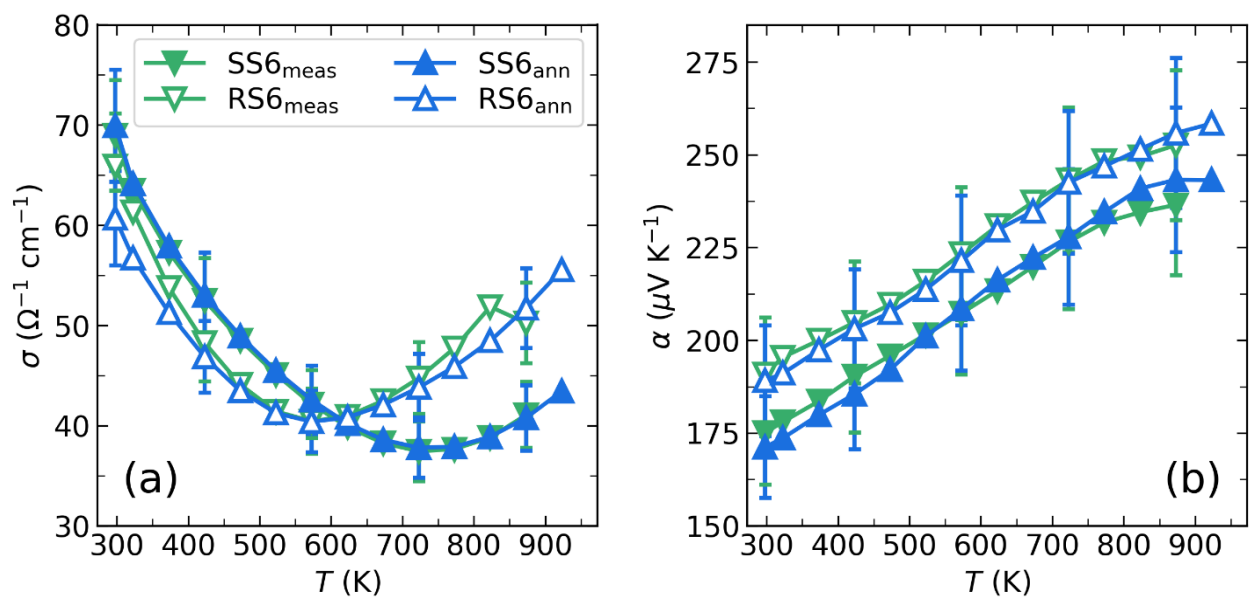

**Figure S5.** Temperature dependence of (a) the electrical conductivity, $\sigma$, (b) the Seebeck coefficient, $\alpha$, for the $Bi_{0.94}Sm_{0.06}CuSeO$ sample prepared by SS and RS methods (solid and open symbols, respectively).

**Details on XPS analysis**

After careful literature review, we think that in the case of Sm-doped BiCuSeO the XPS analysis is not reliable. The XPS spectrum of Sm exhibits a complex multidoublets structure, which is common for all lanthanoids [14]. In some reports, a hyperfine splitting of each Sm 3$d$ peak or the presence of shoulders were observed and attributed to the presence of $Sm^{2+}$ or $Sm^0$ [15–17]. The relative proportions of $Sm^0$, $Sm^{2+}$, and $Sm^{3+}$ are then discussed according to the relative intensity of their apparent peaks. Although spin-orbit splitting observed in this region can, in principle, be related to the ratio of $Sm^{2+}/Sm^{3+}$, the degree of splitting is insufficient for a reliable deconvolution procedure to be attempted. Moreover, there are conflicting reports concerning whether the peaks ascribed to $Sm^{2+}$ appear at higher or lower binding energy relative to the $Sm^{3+}$ doublet depending upon the properties of the material further draws the reliability of such analyses into question. In a previous paper on Sm-doped BiCuSeO [18] the satellite peak at 1070 – 1080 eV was attributed to the $Sm^{2+}$ and this was wrong. Moreover, in the case of Sm-doped BiCuSeO the Cu 2$s$ peak overlaps with a potential peak from $Sm^{2+}$ (see Fig. S5g and Refs. [17,19]). All this means that this form of interpretation can be misleading as it requires extensive peak deconvolution incorporating arbitrary constraints on fitting parameters; in other words, it is possible to get any desirable result, which is mostly attributed to the parameters used for the peak fitting but not the properties of the sample (see Fig. S5f and Fig. S5g). Unfortunately, in this case, the core level O 1$s$ XPS spectrum could not provide any additional insight regarding the presence or absence of $Sm^{2+}$.

In order to support this discussion, we also carried out the XPS for $Bi_{0.92}Sm_{0.08}CuSeO$ prepared by the SS technique. As was already shown in several reports BiCuSeO is not a typical ionic compound and thus, the oxidation states of the constituent elements are not explicitly defined [13,20–22]. In short, all the obtained results for Bi, Cu, Se and, O are in good agreement with previous reports (Fig. S5) [13,20–22]. Most of Bi is in the 3+ state with some amount of $Bi^{2+}$ presented (Fig. S5b). Such an observation was also reported in previous reports suggesting that the change in the Bi oxidation state is relatively easy due to the small energy difference between various oxidation states. According to the XPS spectrum around the binding energy of Cu 2$p$ displayed in Fig. S5c, the Cu is presented in $Cu^+$, $Cu^{2+}$, and $Cu^{3+}$ states. So, the charge compensation may be realized between $Bi^{2+}$ and $Cu^{2+}/Cu^{3+}$. The XPS spectrum of Se 3$d$ is fitted by three doublets (Fig. S5d), first two doublets correspond mainly to the Se–Cu bond with Se in 2– state, a doublet at the range of 58 – 61 eV suggests that there may exist some degree of bonding between Se–O and thus Se oxidation (e.g. $SeO_2$). The wide peak at the binding energy of 528 – 534 eV corresponds to the oxygen chemical state of 2– and 1– (Fig. S5e). The spectrum of Sm 3$d$ is represented by the spin-orbit doublets of 3$d_{5/2}$ and 3$d_{3/2}$ with an energy splitting of 26.8 eV and an integral intensity ratio of 0.67, which is in good agreement with previous reports [23–25] and this can be reliably assigned to $Sm^{3+}$ (see Fig. R1f). At the same time, the deconvolution of the experimental Sm 3$d$ peaks can be also performed considering both $Sm^{3+}$ and $Sm^{2+}$ states with more or less reliable fitting (Fig. R1g) as discussed above. Therefore, we believe that within the framework of this work it is impossible to reliably establish the $Sm^{2+}/Sm^{3+}$ ratio in the samples. Thus, we have to admit, that all discussions related to the presence of $Sm^{2+}$ and its influence on structural and transport properties are speculative. At the same time, to our understanding, the absence of $Sm^{2+}$ ions in the system cannot be unambiguously confirmed either.

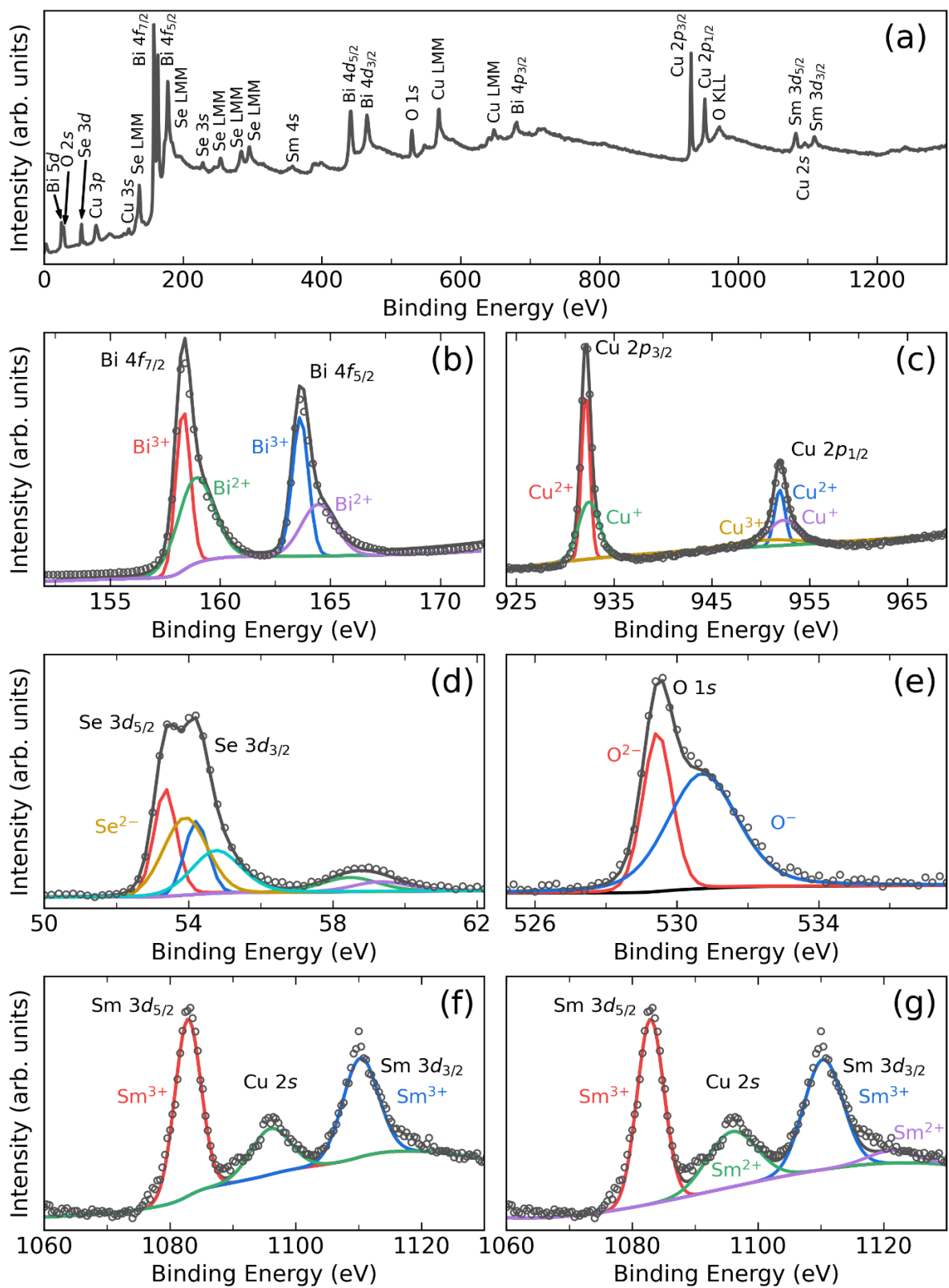

**Figure S6.** (a) XPS survey spectrum of the Bi$_{0.92}$Sm$_{0.08}$CuSeO prepared by SS technique and corresponding high resolution XPS spectra at the binding energy for (b) Bi 4*f*, (c) Cu 2*p*, (d) Se 3*d*, (e) O 1*s*, (f) Sm 3*d* considering only Sm$^{3+}$ and (g) Sm 3*d* considering both Sm$^{3+}$ and Sm$^{2+}$. Open circles are the measured experiment points, solid lines are the background, peak fittings and their sum. The peak positions of different oxidation states are referred to Refs. [13,17,19] and NIST X-ray Photoelectron Spectroscopy Database [26].

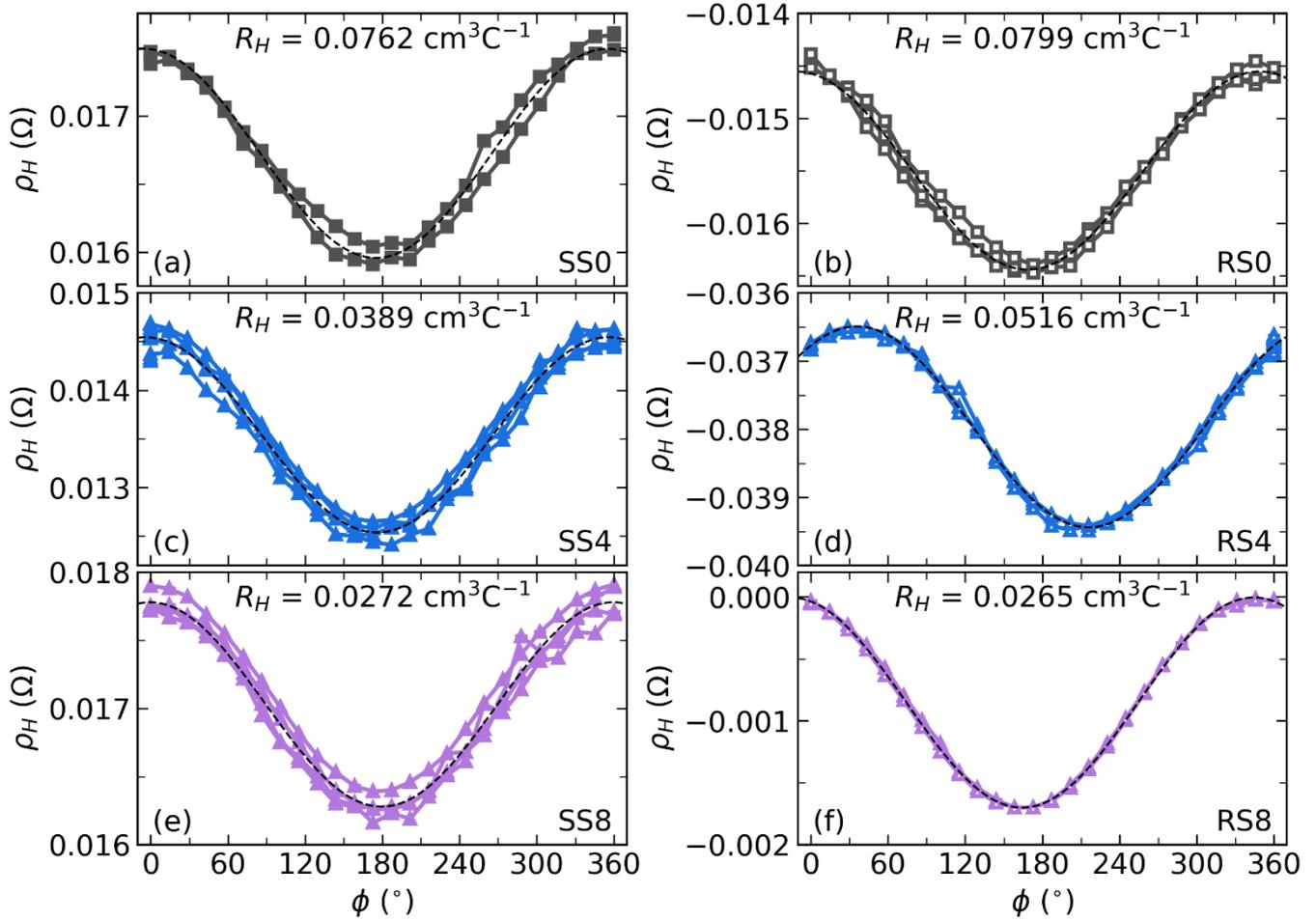

**Figure S7.** The experimental angular dependence of Hall resistivity and its cosine fit (black dashed lines) for the Bi$_{1-x}$Sm$_x$CuSeO ($x$ = 0, 0.04, 0.08) samples prepared by (a, c, e) SS and (b, d, f) RS methods representing good reproducibility for several runs of measurements.

**Table S4.** Room-temperature Hall carrier concentration and Hall carrier mobility of Bi$_{1-x}$Sm$_x$CuSeO ($x$ = 0, 0.04, 0.08) samples prepared by SS and RS methods; values reported by Feng *et al.* [27] and Kang *et al.* [18] are also presented for comparison.

| Nominal composition | $p_H$ (cm$^{-3}$) | | | | $\mu_H$ (cm$^2$ V$^{-1}$ s$^{-1}$) | | | |
|---|---|---|---|---|---|---|---|---|
| | SS | RS | Feng *et al.* [27] | Kang *et al.* [18] | SS | RS | Feng *et al.* [27] | Kang *et al.* [18] |
| BiCuSeO | 8.20·10$^{19}$ | 7.82·10$^{19}$ | 1.5·10$^{19}$ | 8.96·10$^{18}$ | 3.41 | 3.69 | 6.77 | 2.30 |
| Bi$_{0.975}$Sm$_{0.025}$CuSeO | n/a | n/a | n/a | 4.29·10$^{19}$ | n/a | n/a | n/a | 2.27 |
| Bi$_{0.96}$Sm$_{0.04}$CuSeO | 1.61·10$^{20}$ | 1.21·10$^{20}$ | 1.69·10$^{19}$ | n/a | 2.38 | 2.80 | 12.55 | n/a |
| Bi$_{0.95}$Sm$_{0.05}$CuSeO | n/a | n/a | n/a | 6.80·10$^{19}$ | n/a | n/a | n/a | 2.41 |
| Bi$_{0.94}$Sm$_{0.08}$CuSeO | 2.29·10$^{20}$ | 2.36·10$^{20}$ | 1.88·10$^{19}$ | n/a | 2.23 | 2.07 | 22.58 | n/a |
| Bi$_{0.90}$Sm$_{0.10}$CuSeO | n/a | n/a | n/a | 8.08·10$^{19}$ | n/a | n/a | n/a | 4.39 |

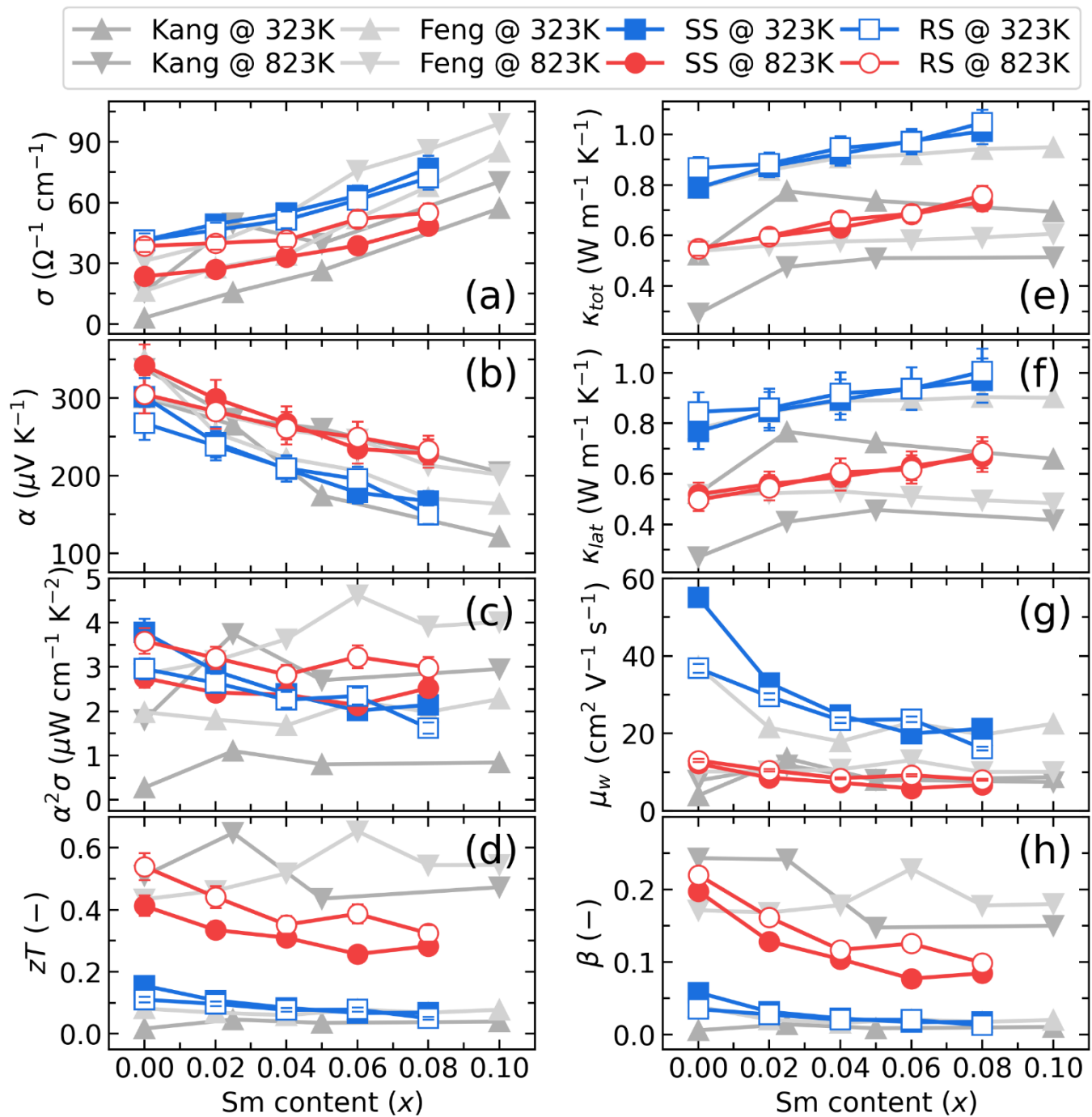

**Figure S8.** (a) Electrical conductivity, $\sigma$, (b) Seebeck coefficient, $\alpha$, (c) power factor, $\alpha^2\sigma$, (d) figure of merit, $zT$, (e) total thermal conductivity, $\kappa_{tot}$, (f) lattice thermal conductivity, $\kappa_{lat}$, (g) weighted mobility, $\mu_w$, (h) quality factor, $\beta$, as a function of doping fraction on Bi site at 323 K (blue) and 823 K (red) for the $Bi_{1-x}Sm_xCuSeO$ ($x = 0 – 0.08$) samples prepared by SS and RS methods (solid and open symbols, respectively) with additional data from the literature [18,27].

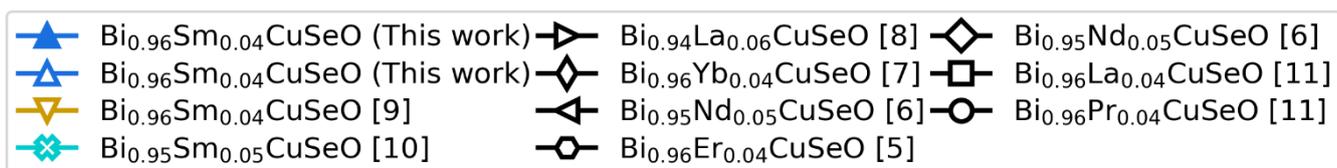
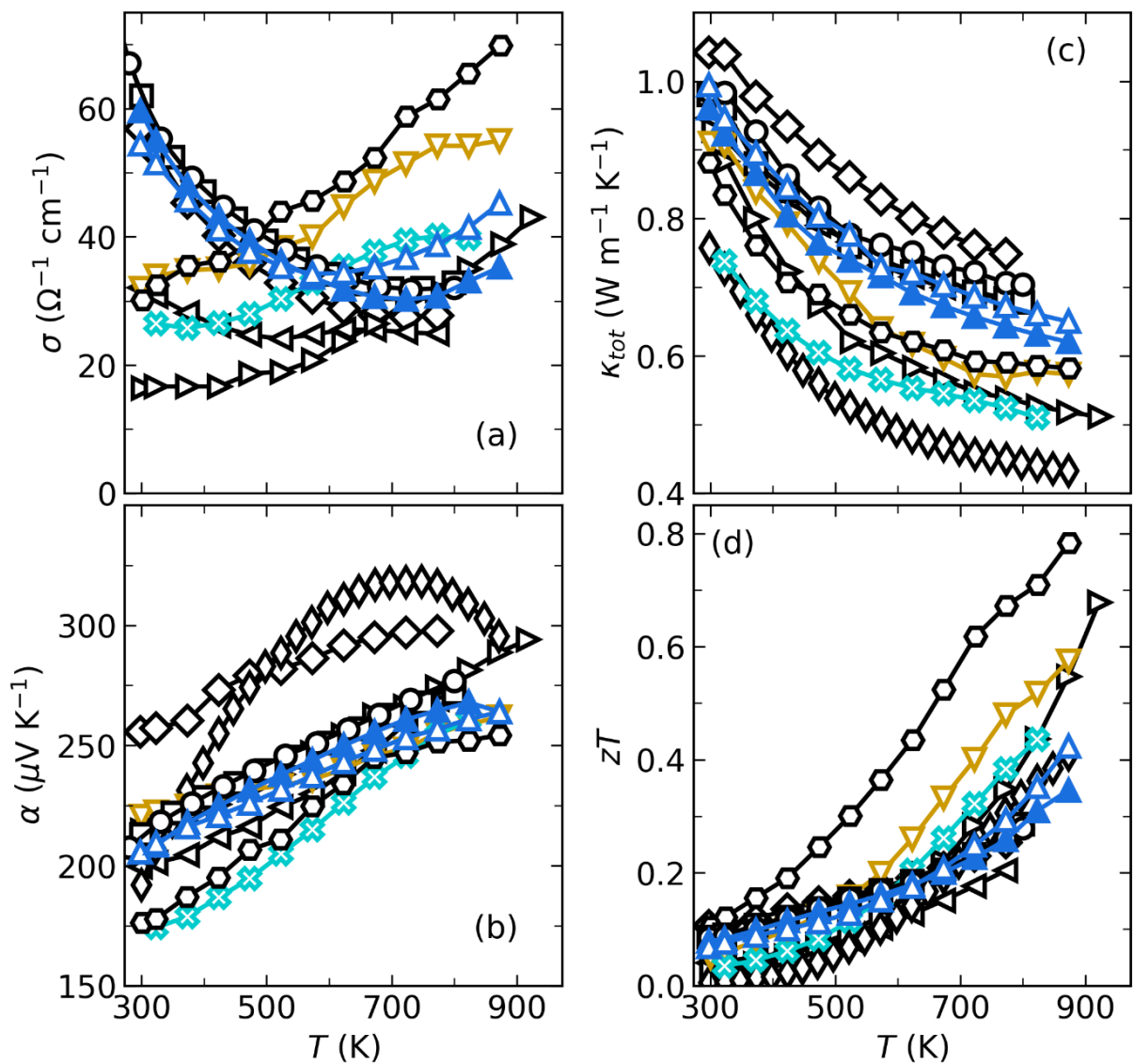

**Figure S9.** (a) Electrical conductivity, $\sigma$, (b) Seebeck coefficient, $\alpha$, (c) total thermal conductivity, $\kappa_{tot}$, (d) figure of merit, $zT$, for the $Bi_{0.96}Sm_{0.04}CuSeO$ obtained in this work (blue triangles; filled symbols for SS route, empty symbols for RS route), $Bi_{0.96}Sm_{0.04}CuSeO$ reported by Feng *et al.* [27], and $Bi_{0.95}Sm_{0.05}CuSeO$ reported by Kang *et al.* [18]. Data reported for other rare-earth doped BiCuSeO are also presented for comparison [28–32].

This work, SS series
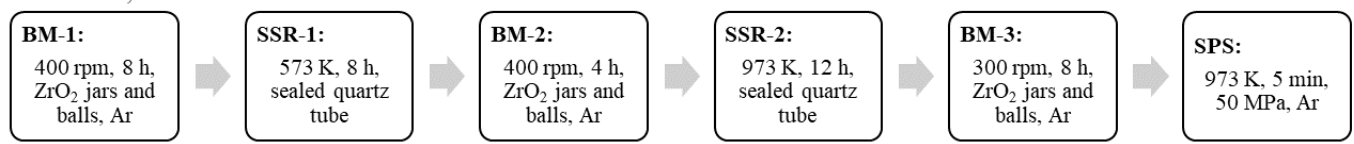

This work, RS series
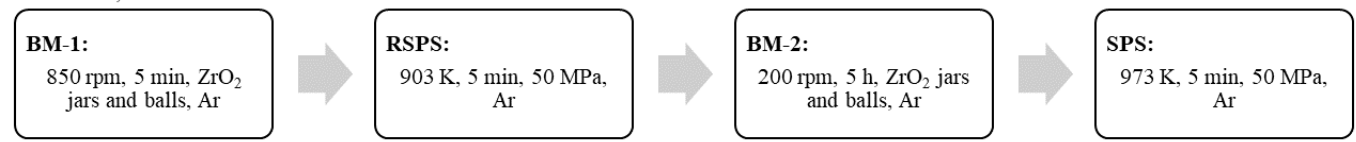

Feng *et al.*
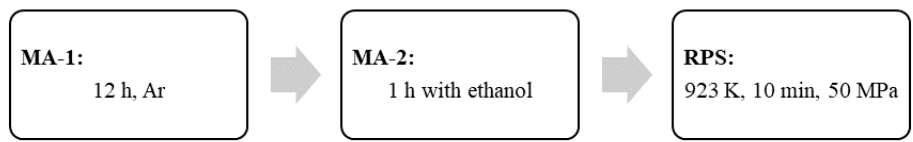

Kang *et al.*
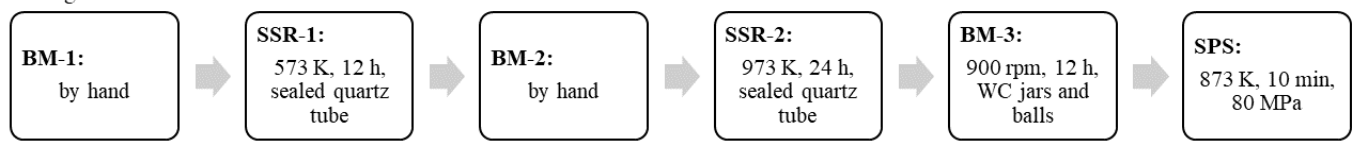

**Figure S10.** Schematic illustration of fabrication routes used in this work and work reported by Feng *et al.* [27], and Kang *et al.* [18]. Here BM is the ball milling step (mixing, grinding, etc.), SSR is the solid-state reaction step, SPS is the spark plasma sintering and MS is the mechanical alloying.

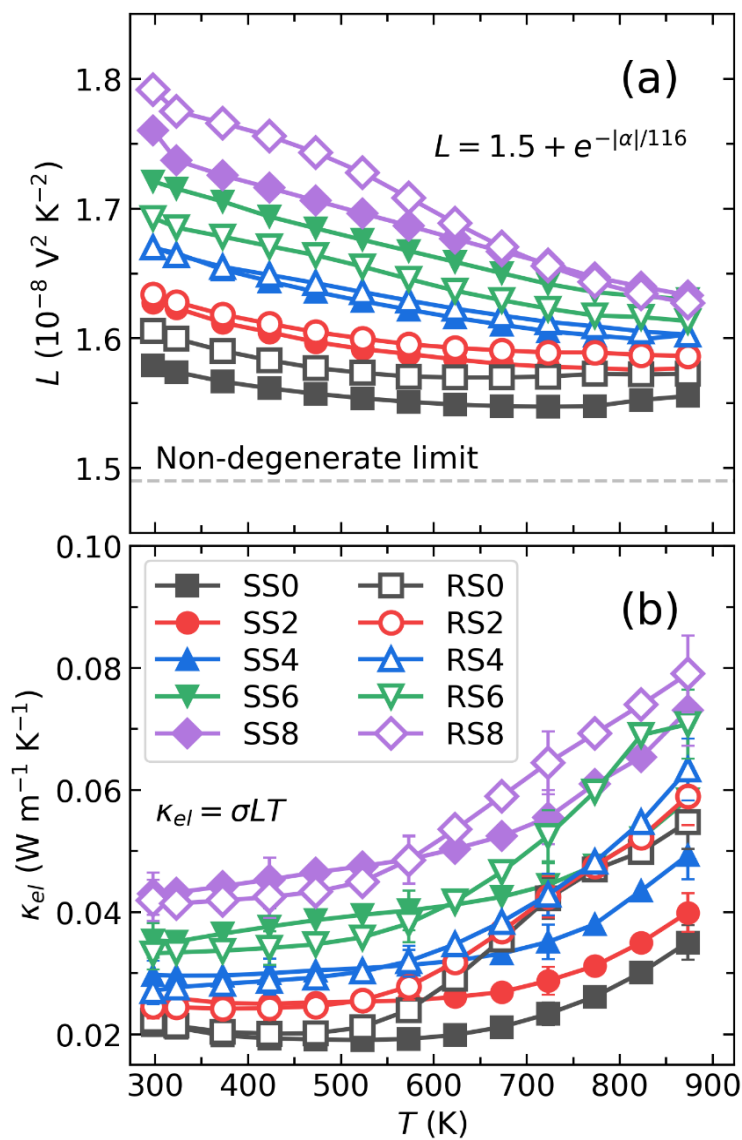

**Figure S11.** Temperature dependence of (a) the Lorenz number, *L*, and (b) the electronic thermal conductivity, $\kappa_{el}$, for the $Bi_{1-x}Sm_xCuSeO$ ($x$ = 0 – 0.08) samples prepared by SS and RS methods (solid and open symbols, respectively).

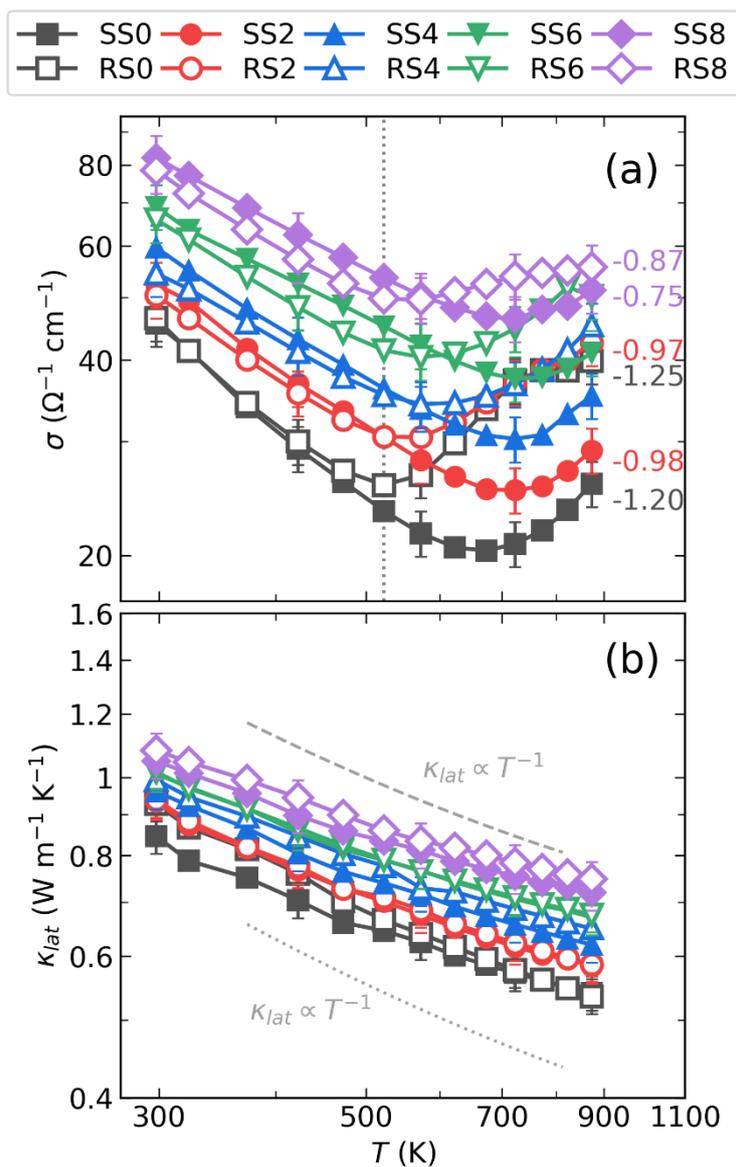

**Figure S12.** Log-log scale for the temperature dependence of (a) the electrical conductivity and (b) the lattice thermal conductivity for the $Bi_{1-x}Sm_xCuSeO$ ($x = 0 - 0.08$) samples prepared by SS and RS methods (solid and open symbols, respectively). The scattering factor for some samples is indicated to the right of the curves of the corresponding color.

**Table S5.** Average mass and sound velocity, volume per atom, the number of atoms per primitive cell and Grüneisen parameter for the BiCuSeO and SmCuSeO.

| Material | $\bar{M}$ (g mol$^{-1}$) | $v_a$ (m s$^{-1}$) | $V_{at}$ (Å$^3$) | $n_{at}$ (–) | $\gamma$ |
|---|---|---|---|---|---|
| BiCuSeO | 91.87 | 2107 [33] | 17.15 | 8 | 1.5 [33] |
| SmCuSeO | 77.22 | 2561 [34] | 17.05 [7] | 8 | ~1.2 |

**Table S6.** Electronegativity for O, Bi and Sm. Comparison of electronegativity difference and Pauling ionicity for the Bi − O and Sm − O bonds, respectively.[*]

| Property | Sm | Bi | O | Bi – O | Sm – O |
|---|---|---|---|---|---|
| Electronegativity, $\chi$ | 1.9 | 2.69 | 3.78 | n/a | n/a |
| Electronegativity difference, $\Delta\chi$ | n/a | n/a | n/a | 1.09 | 1.88 |
| Pauling ionicity, $f$ | n/a | n/a | n/a | 0.55 | 0.91 |

[*]All the electronegativity values were taken from [35].

**References**

[1]   O. Madelung, G.K. White, eds., Thermal Conductivity of Pure Metals and Alloys, Springer-Verlag, Berlin/Heidelberg, 1991.

[2]   B. Lizell, J. Chem. Phys. 20 (1952) 672–676.

[3]   S. Iyyapushpam, P. Chithra lekha, D. Pathinettam Padiyan, Phys. B Condens. Matter 405 (2010) 712–719.

[4]   O. Madelung, U. Rössler, M. Schulz, eds., Non-Tetrahedrally Bonded Elements and Binary Compounds I, Springer-Verlag, Berlin/Heidelberg, 1998.

[5]   G.V.S. Rao, S. Ramdas, P.N. Mehrotra, C.N.R. Rao, J. Solid State Chem. 2 (1970) 377–384.

[6]   W.M. Haynes, ed., CRC Handbook of Chemistry and Physics, 97th ed., CRC Press, Taylor & Francis Group, Boca Raton, 2017.

[7]   T. Ohtani, M. Hirose, T. Sato, K. Nagaoka, M. Iwabe, Jpn. J. Appl. Phys. 32 (1993) 316–318.

[8]   W.J. Zhu, Y.Z. Huang, C. Dong, Z.X. Zhao, Mater. Res. Bull. 29 (1994) 143–147.

[9]   A.M. Kusainova, P.S. Berdonosov, L.G. Akselrud, L.N. Kholodkovskaya, V.A. Dolgikh, B.A. Popovkin, J. Solid State Chem. 112 (1994) 189–191.

[10]  P.S. Berdonosov, A.M. Kusainova, L.N. Kholodkovskaya, V.A. Dolgikh, L.G. Akselrud, B.A. Popovkin, J. Solid State Chem. 118 (1995) 74–77.

[11]  Y. Ohki, S. Komatsuzaki, Y. Takahashi, K. Takase, Y. Takano, K. Sekizawa, in: AIP Conf. Proc., AIP, 2006, pp. 1309–1310.

[12]  C. Barreteau, D. Bérardan, L. Zhao, N. Dragoe, J. Mater. Chem. A 1 (2013) 2921–2926.

[13]  C.-L. Hsiao, X. Qi, Acta Mater. 102 (2016) 88–96.

[14]  D.D. Sarma, M.S. Hegde, C.N.R. Rao, J. Chem. Soc. Faraday Trans. 2 77 (1981) 1509.

[15]  B.T. Sone, E. Manikandan, A. Gurib-Fakim, M. Maaza, J. Alloys Compd. 650 (2015) 357–362.

[16]  Q. Liu, H. Yang, H. Dong, W. Zhang, B. Bian, Q. He, J. Yang, X. Meng, Z. Tian, G. Zhao, New J. Chem. 42 (2018) 13096–13106.

[17]  T.-D. Nguyen, D. Mrabet, T.-O. Do, J. Phys. Chem. C 112 (2008) 15226–15235.

[18]  H. Kang, X. Zhang, Y. Wang, J. Li, D. Liu, Z. Chen, E. Guo, X. Jiang, T. Wang, Mater. Res. Bull. 126 (2020) 110841.

[19]  G.K. Wertheim, G. Crecelius, Phys. Rev. Lett. 40 (1978) 813–816.

[20]  D.H. Kim, H.Y. Hong, J.K. Lee, S.D. Park, K. Park, J. Mater. Res. Technol. 9 (2020) 16202–16213.

[21]  M. Ishizawa, Y. Yasuzato, H. Fujishiro, T. Naito, H. Katsui, T. Goto, J. Appl. Phys. 123 (2018) 245104.

[22]  D. Yuan, S. Guo, S. Hou, Y. Ma, J. Wang, S. Wang, Nanoscale Res. Lett. 13 (2018) 382.

[23]  D. Duan, C. Hao, W. Shi, H. Wang, Z. Sun, RSC Adv. 8 (2018) 11289–11295.

[24]  S. Biswas, H. Naskar, S. Pradhan, Y. Chen, Y. Wang, R. Bandyopadhyay, P. Pramanik, New J. Chem. 44 (2020) 1921–1930.

[25]  H. Yu, J. Xu, H. Guo, Y. Li, Z. Liu, Z. Jin, RSC Adv. 7 (2017) 56417–56425.

[26]  (n.d.).

[27]  B. Feng, X. Jiang, Z. Pan, L. Hu, X. Hu, P. Liu, Y. Ren, G. Li, Y. Li, X. Fan, Mater. Des. 185 (2020) 108263.

[28]  Y. Liu, J. Ding, B. Xu, J. Lan, Y. Zheng, B. Zhan, B. Zhang, Y. Lin, C. Nan, Appl. Phys. Lett. 106 (2015) 233903.

[29]  H. Kang, J. Li, Y. Liu, E. Guo, Z. Chen, D. Liu, G. Fan, Y. Zhang, X. Jiang, T. Wang, J. Mater. Chem. C 6 (2018) 8479–8487.

[30]  A. Novitskii, G. Guélou, D. Moskovskikh, A. Voronin, E. Zakharova, L. Shvanskaya, A. Bogach, A. Vasiliev, V.

Khovaylo, T. Mori, J. Alloys Compd. 785 (2019) 96–104.

[31]  B. Feng, G. Li, Z. Pan, X. Hu, P. Liu, Y. Li, Z. He, X. Fan, Ceram. Int. 45 (2019) 4493–4498.

[32]  A. Novitskii, I. Serhiienko, S. Novikov, Y. Ashim, M. Zheleznyi, K. Kuskov, D. Pankratova, P. Konstantinov, A. Voronin, O. Tretiakov, T. Inerbaev, A. Burkov, V. Khovaylo, (2021) http://arxiv.org/abs/2104.10509.

[33]  L.-D. Zhao, J. He, D. Berardan, Y. Lin, J.-F. Li, C. Nan, N. Dragoe, Energy Environ. Sci. 7 (2014) 2900–2924.

[34]  J. He, Y. Xia, W. Lin, K. Pal, Y. Zhu, M.G. Kanatzidis, C. Wolverton, (2021) arXiv:2107.04955.

[35]  C. Tantardini, A.R. Oganov, Nat. Commun. 12 (2021) 2087.